\documentclass[12pt]{article}

\usepackage{epsfig} 
\usepackage{amsbsy}

\def\beq{\begin{equation}}
\def\eeq{\end{equation}}

\def\beqn{\begin{eqnarray}}
\def\eeqn{\end{eqnarray}}

\def\({\left(}
\def\){\right)}

\def\Dx{\int {\cal D}x }
\def\Q{{\cal Q} }
\def\P{{\cal P} }
\def\M{{\cal M} }
\def\Pre{{\cal PRE} }
\def\Sum{{\cal SUM} }

\newcommand{\G}[1]{\Gamma\left({#1}\right)}

\newcommand{\halfD}{\frac{D}{2}}
\newcommand{\halfs}{\frac{\sigma}{2}}

\def\Id{I^D_n\(\{\nu_i\};\{Q_i^2\},\{M_i^2\}\)}
\def\Ideq{I^D_n\(\{\nu_i\};\{Q_i^2\},M^2\)}
\def\poch{Pochhammer}

\def\f21{{_2F_1}}
\def\ftt{{_3F_2}}

\def\al{\alpha}
\def\ap{\alpha'}
\def\bp{\beta'}
\def\bt{\beta}
\def\ga{\gamma}
\def\gp{\gamma'}
\def\de{\delta}
\def\dl#1{$$\displaylines{\quad#1}$$}

\def\li#1{\,{\rm Li}_2\left(#1 \right)}
\def\Re{\mathop{\rm Re}}

\catcode`\@=11

\@addtoreset{equation}{section}

\def\theequation{\thesection.\arabic{equation}}

\def\@normalsize{\@setsize\normalsize{15pt}\xiipt\@xiipt
\abovedisplayskip 14pt plus3pt minus3pt%
\belowdisplayskip \abovedisplayskip
\abovedisplayshortskip \z@ plus3pt%
\belowdisplayshortskip 7pt plus3.5pt minus0pt}

\def\small{\@setsize\small{13.6pt}\xipt\@xipt
\abovedisplayskip 13pt plus3pt minus3pt%
\belowdisplayskip \abovedisplayskip
\abovedisplayshortskip \z@ plus3pt%
\belowdisplayshortskip 7pt plus3.5pt minus0pt
\def\@listi{\parsep 4.5pt plus 2pt minus 1pt
     \itemsep \parsep
     \topsep 9pt plus 3pt minus 3pt}}

\@twosidetrue

\catcode`@=12

\catcode`\@=11

\def\section{\@startsection{section}{1}{\z@}{3.5ex plus 1ex minus
   .2ex}{2.3ex plus .2ex}{\large\bf}}

\def\thesection{\arabic{section}}
\def\thesubsection{\arabic{section}.\arabic{subsection}}
\def\thesubsubsection{\arabic{section}.\arabic{subsection}.\arabic{subsubsection}}

\def\appendix{\setcounter{section}{0}
 \def\thesection{\Alph{section}}
 \def\theequation{\Alph{section}.\arabic{equation}}
 \def\thesubsection{\Alph{section}.\arabic{subsection}}
\def\thesubsubsection{\Alph{section}.\arabic{subsection}.\arabic{subsubsection}}

 \def\section{\@startsection{section}{1}{\z@}{3.5ex plus 1ex minus
   .2ex}{2.3ex plus .2ex}{\large\bf}}
}

\def \ep{\epsilon}
\def \to   {\mbox{$\rightarrow$}}

\newcommand\hepph[1]{{\tt hep-ph/#1}}
\newcommand\hepth[1]{{\tt hep-th/#1}}

\def\ord#1{{\cal O}\(#1\)}

\begin{document}
\begin{titlepage}
\nopagebreak
{\flushright{
        \begin{minipage}{4cm}
         DTP/99/80 \\
        {\tt hep-ph/9907494}\hfill \\
        \end{minipage}        }

}
\vfill
\begin{center}
{\LARGE \bf \sc
 \baselineskip 0.9cm
Scalar One-Loop Integrals using the Negative-Dimension Approach

 }
\vskip
1.3cm 
{\large  C.~Anastasiou\footnote{e-mail: {\tt Ch.Anastasiou@durham.ac.uk}},
E.~W.~N.~Glover\footnote{e-mail: {\tt E.W.N.Glover@durham.ac.uk}}  and
C.~Oleari\footnote{e-mail: {\tt Carlo.Oleari@durham.ac.uk}}} 
\vskip .2cm 
{\it Department of Physics, 
University of Durham, 
Durham DH1 3LE, 
England } 
\vskip
1.3cm    
\end{center}

\nopagebreak
%\vfill
%\vskip 3cm
\begin{abstract}

We study massive one-loop integrals by analytically continuing the Feynman
integral to negative dimensions as advocated by Halliday and Ricotta and
developed by Suzuki and Schmidt.  We
consider $n$-point one-loop integrals with arbitrary powers of propagators in
general dimension $D$. For integrals with $m$ mass scales and $q$ external
momentum scales, we construct a template solution valid for all $n$ which
allows us to obtain a representation of the graph in terms of a finite sum of
generalised hypergeometric functions with $m+q-1$ variables.  All solutions
for all possible kinematic regions are given simultaneously, allowing the
investigation of different ranges of variation of mass and momentum scales.

As a first step, we develop the general framework and apply it to massive
bubble and vertex integrals.  Of course many of these integrals are well
known and we show that the known results are recovered.  To give a concrete
new result, we present expressions for the general vertex integral with one
off-shell leg and two internal masses in terms of hypergeometric functions of
two variables that converge in the appropriate kinematic regions.  The
kinematic singularity structure of this graph is sufficiently complex to give
insight into how the negative-dimension method operates and gives some hope
that more complicated graphs can also be evaluated.

\end{abstract}
\vfill
%\today \timestamp \hfill
\vfill
\end{titlepage}
\newpage                                                                     
%%%%%%%%%%%%%%%%%%%%%%%%%%%%%%%%%%%%%%%%%%%%%%%%% 
\begin{center}
  \parbox{.95\textwidth}{
      {
%\footnotesize
%\small
\normalsize
        \tableofcontents}
    }
\end{center}

%%%%%%%%%%%%%%%%%%%%%%%%%%%%%%%%%%%%%%%%%%%%%%%%% 
\newpage

%%%%%%%%%%%%%%%%%%%%%%%%%%%%%%%%%%%%%%%%%%%%%%%%% 
\section{Introduction} 
\setcounter{equation}{0}
\label{sec:intro}
 
Loop integrals play an important role in making precise perturbative
predictions in quantum field theory, in general, and in the Standard Model
of particle physics, in particular.   As such, a large effort has been expended
in developing methods for evaluating them. The problem is complicated by the
appearance of ultraviolet (UV) and infrared (IR) singularities, and it has
become customary to use dimensional regularisation~\cite{DR,HV,Collins} to
extend the dimensionality of the loop integral away from 4-dimensions to
$D=4-2\epsilon$, to regulate the infrared and ultraviolet singularities.  

With the increasing of the number of legs, of the number of mass scales or of
the number of loops, the integrals can be made almost arbitrarily complex and
difficult to solve analytically. Different
%methods~\cite{HV2,CT,Terrano,ibyp,uniq,Jos,kotikov,BDK} 
methods~\cite{HV2}--\cite{BDK} 
have been developed to solve the Feynman integrals. We mention here only two
of them: the integration by parts~\cite{ibyp}, which works well for some
two-loop vertex diagrams~\cite{KL} reducing them to simpler known graphs with
different powers of the propagators (however this breaks down for more
complicated graphs such as the double box, where irreducible numerators
factors are present) and the Mellin-Barnes integral representation (see for
example~\cite{BD}), which was successfully used by Smirnov~\cite{Smirnov} to
calculate the two-loop box integral.  In this approach, the integral is
usually written as multiple contour integrals of $\Gamma$ functions and
powers of ratios of the mass scales in the problem. By closing the contour,
we obtain an infinite series of residues at the singular points of the
$\Gamma$ functions.  These series can be identified as generalised
hypergeometric functions, whose convergence properties reflect the
threshold-singularity structure of the integral.

There are several advantages in using hypergeometric functions to represent
the integral.  First, these hypergeometric functions often have integral
representations themselves, in which an expansion in $\epsilon$ can be made,
yielding expressions in logarithms, dilogarithms etc..  It seems that, where
direct evaluation of the hypergeometric function in terms of known functions
is possible, very compact results are obtained~\cite{Dbox, Sanchis}.  Second,
because the series is convergent and well behaved in a particular region of
phase space, it can be numerically evaluated~\cite{Frits}.  In fact, each
hypergeometric representation immediately allows an asymptotic expansion of
the integral in terms of ratios of momentum and mass scales.  Third, through
analytic continuation formulae, the hypergeometric functions valid in one
kinematic domain can be re-expressed in a different kinematic region.

Not all work has concentrated around $D=4$. In fact, a close connection
between tensor loop integrals - those with additional powers of the loop
momentum in the numerator - and higher-dimension scalar integrals
($D=6-2\epsilon$, for example) is well established~\cite{BDK, CGM, Tar1,
Tar2}.  Furthermore, in 1987, Halliday and Ricotta~\cite{HR,DH} suggested
that it would be useful to calculate the loop integral considering $D$ as a
negative number.  Because loop integrals are analytic in the number of
dimensions $D$ (and also in the powers of the propagators) they proposed to
calculate the integral in negative dimensions and return to positive
dimensions, and specifically $D=4-2\epsilon$, after the integrations have
been performed.  As we will discuss more fully later on, integration over the
loop momentum and/or the parameters introduced to do the loop integration is
replaced with infinite series, which again can be identified as generalised
hypergeometric functions.  Recently this idea has been picked up again by
Suzuki and Schmidt who have evaluated a number of two-loop
integrals~\cite{SS2loop}, three-loop integrals~\cite{SS3loop}, one-loop
tensor integrals~\cite{SStensor} as well as the one-loop massive box integral
for the scattering of light by light~\cite{SSbox}.  In this latter case, as
well as reproducing the known hypergeometric-series representations
of Ref.~\cite{Dbox}, valid in particular kinematic regions, Suzuki and Schmidt
simultaneously found hypergeometric solutions valid in other kinematic
domains.  Of course, all of these solutions are related by analytic
continuation.  However, it is easy to envisage integrals that yield
hypergeometric functions where the analytic continuation formulae are not
known a priori.  In these cases, having series expansions directly available
in all kinematic regions is useful.

In this paper we wish to explore the negative-dimension approach (NDIM)
further. In particular we focus on one-loop integrals with general powers of
the propagators and arbitrary dimension $D$.  There are several reasons for
doing this.  First, it allows connection with the general tensor-reduction
program based on integration by parts of Refs.~\cite{Tar1, Tar2}.  Here the
tensor integrals are linear combinations of scalar integrals with either
higher dimension or propagators raised to higher powers. Second, we can
imagine inserting the one-loop results into a two-loop integral by closing up
external legs.  This is trivial for most bubble integrals, but more
complicated for vertex and box graphs.  Broadhurst~\cite{broad} has shown
that this is possible for the non-trivial two-loop self-energy graph. Third,
it actually simplifies the calculation.  As we will show, by keeping the
parameters general, it is easier to identify the regions of convergence of the
hypergeometric series and therefore which hypergeometric functions to group
together. For specific values of the parameters, the hypergeometric functions
often collapse to simpler functions.  As a first step, we develop the general
framework and apply it to massive bubble and vertex integrals.  Of course
many of these results are well known.  However, they serve to iron out some
of the subtleties of the NDIM approach.  To give a concrete new result, we
present expressions for the vertex integral with one off-shell leg and two
internal masses in arbitrary dimension and for general powers of the
propagators in terms of hypergeometric functions of two variables.  The
kinematic singularity structure of this graph is sufficiently complex to give
insight into how NDIM operates and gives some hope that more complicated
graphs can also be evaluated.

Our paper is organised as follows.  In Sec.~\ref{sec:general} we first review
the theoretical framework of one-loop integrals with Schwinger parameters and
briefly explain the basic idea of integrating in negative dimensions.  We
then apply NDIM to construct template solutions for arbitrary one-loop
integrals together with a linear system of constraints that relates the
powers of the propagators in the loop integral to the summation variables.
The system of constraints has many solutions and each one must be inserted
into the template solution, yielding a sum over fewer variables that can be
identified as a generalised hypergeometric function.
This method gives simultaneously all the solutions in all the possible
kinematic regions. 
The approach is illustrated for the massive bubble integral where we show how
to recover the known results.  We discuss how the form of the solution in
different kinematical regimes is dictated by the convergence properties of
the hypergeometric functions and the structure of the system.  In
Sec.~\ref{sec:triangles} we consider one-loop triangles and give the form of
the template solution and the system of constraints  with
arbitrary powers of the propagators, internal masses, external legs off-shell
and for general $D$.  We apply this result to triangle integrals with three
scales and give expressions valid in the various kinematic regions
appropriate to the vertex integral.  Results are given in the form of
hypergeometric functions of one and two variables, which are defined in 
Appendix~\ref{sec:app}.  For specific choices of $D$ and the propagator
powers, these 
functions can be evaluated as logarithms and dilogarithms using the integral
representations that are also provided in the appendix together with a list
of hypergeometric identities that often simplify the results.  Finally, our
method is summarized in Sec.~\ref{sec:conc}.

%%%%%%%%%%%%%%%%%%%%%%%%%%%%%%%%%%%%%%%%%%%%%%%%% 
%\section{The negative-dimension method} 
%\setcounter{equation}{0}
%%%%%%%%%%%%%%%%%%%%%%%%%%%%%%%%%%%%%%%%%%%%%%%%% 

\section{Theoretical framework}

\begin{figure}
\begin{center}
~\epsfig{file=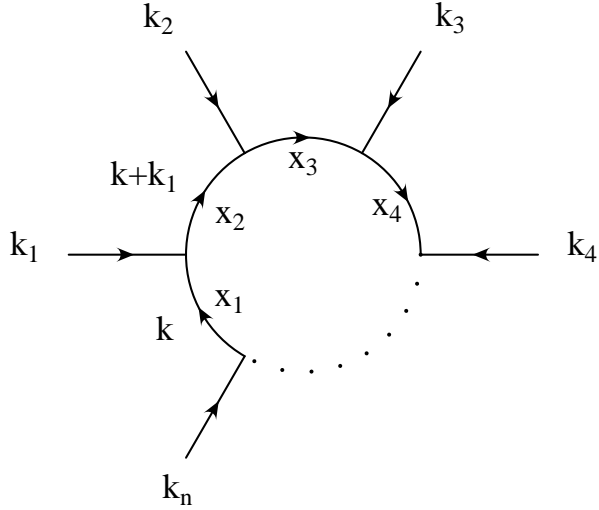,height=7cm}
\end{center}
\caption[]{The generic one-loop graph}
\label{fig:generic}
\end{figure}

The generic $n$-point one-loop integral in $D$-dimensional Minkowski
space with loop momentum $k$ is given by
\begin{equation} 
\label{eq:IDloop}
\Id = 
\int  \frac{d^Dk}{i\pi^{D/2}} \frac{1}{A_1^{\nu_1}\ldots A_n^{\nu_n}}, 
\end{equation} 
where, as indicated in Fig.~\ref{fig:generic},
the external momenta $k_i$ are all incoming so that $\sum_{i=1}^n
k_i^\mu = 0$ and the propagators have the form
\begin{eqnarray} 
A_1 &=& k^2 -M_1^2 + i0, \nonumber \\
A_i &=& \left(k+\sum_{j=1}^{i-1} k_j\right)^2 -M_i^2 + i0\qquad {i \neq 1},
\end{eqnarray}
$M_i$ being the mass of the $i$th propagator. The external momentum scales
are indicated with $\{Q_i^2\}$.  For standard integrals, the powers $\nu_i$ to
which each propagator is raised are usually unity. However, we wish, where
possible, to leave the powers as general as possible.  As discussed earlier,
this may have some advantages in evaluating two-loop integrals where often
one-loop integrals with arbitrary powers can be inserted into the second loop
integration.

\begin{figure}
\begin{center}
~\epsfig{file=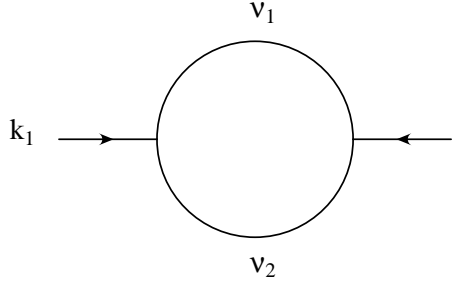,height=4cm}
\end{center}
\caption[]{The one-loop vacuum graph}
\label{fig:bubble}
\end{figure}

To evaluate this integral, we introduce a Schwinger parameter
$x_i$ for each propagator (noting that $A_i < 0$ after Wick rotation 
to Euclidean space) so that
\begin{equation}
\frac{1}{A_i^{\nu_i}} = \frac{(-1)^{\nu_i}}{\Gamma(\nu_i)}
\int^{\infty}_0 dx_i x_i^{\nu_i-1} ~\exp(x_iA_i),
\end{equation}
and we can rewrite Eq.~(\ref{eq:IDloop}) as
\begin{equation}
\Id = \Dx
\int 
\frac{d^Dk}{i\pi^{D/2}}
~\exp\left (\sum_{i=1}^n x_iA_i\right),
\label{eq:form1}
\end{equation}
where we have used the shorthand
\begin{equation}
\Dx =(-1)^\sigma\left(\prod_{i=1}^n \frac{1}{\Gamma(\nu_i)}
\int^{\infty}_0 
dx_i x_i^{\nu_i-1}\right),
\end{equation}
with 
\begin{equation}
\label{eq:sigma}
\sigma = \sum_{i=1}^{n} \nu_i.
\end{equation}
The Gaussian integral over the loop momentum can be solved in a
straightforward way, and using the Minkowski space relation
\begin{equation}
\label{eq:gaussian}
\int \frac{d^Dk}{i\pi^{D/2}} \exp(\alpha k^2) =
\frac{1}{\alpha^{D/2}},
\end{equation}
we have the usual Minkowski space result
\begin{equation}
\Id = \Dx
\frac{1}{\P^{D/2}} ~\exp(\Q/\P)~\exp(-\M).
\label{eq:form2}
\end{equation}
The quantities $\P$ and $\M$ are given by
\begin{eqnarray}
\label{eq:P}
\P &=& \sum_{i=1}^n x_i,  \\
\label{eq:M}
\M &=& \sum_{i=1}^n x_i\,M_i^2,
\end{eqnarray}
while $\Q$ may be simply read off from the Feynman diagram
\begin{equation}
\label{eq:Q}
\Q= \sum_{i=1}^{n-1}\sum_{j=i+1}^n x_i x_j \left(\sum_{k=i}^{j-1}
k_k\right)^2 = \sum_{i=1}^q \Q_i.
\end{equation}
Each of the $q$ terms in $\Q$ is indicated with $\Q_i$, and is obtained by
cutting the loop diagram into two across propagators  
$a$ and $b$ and constructing the four-momentum $Q_i^\mu$ on each side of the
cut: $\Q_i = x_ax_b \,Q_i^2$. 
For example, the one-loop bubble graph 
shown in Fig.~\ref{fig:bubble} has two propagators ($n=2$), so that
$\P = x_1 + x_2$ and $\M =  x_1\,M_1^2 +  x_2\, M_2^2$.
$\Q$ is obtained by examining the momentum flowing across the only possible 
cut ($q=1$): $\Q = \Q_1 = x_1 x_2 \, Q_1^2$, with $Q_1^2 = k_1^2$.

\subsection{The negative-dimension approach}

The crucial point in the negative-dimension approach is that the
Gaussian integral (\ref{eq:gaussian}) is an analytic function of the
space-time dimension.  Hence it is possible to consider $D < 0$ and to
make the definition \cite{HR}
\begin{equation}
\int \frac{d^Dk}{i\pi^{D/2}} \, (k^2)^n = n! \;\delta_{n+\halfD,\, 0}
\label{eq:def}
\end{equation}
for positive values of $n$. 
We see that by expanding the
exponential in (\ref{eq:gaussian}) and inserting the definition
(\ref{eq:def}), after the exchange of the integration with respect to the
summation  
\begin{equation}
\int \frac{d^Dk}{i\pi^{D/2}} \exp(\alpha k^2) =
\sum_{n=0}^{\infty} \frac{\alpha^n}{n!} \int \frac{d^Dk}{i\pi^{D/2}} 
(k^2)^n\nonumber 
= \frac{1}{\alpha^{D/2}},
\end{equation}
we recover the original result, provided that $D$ is both negative and even (so
that the Kronecker $\delta$ can be satisfied and the contribution with $n =
-D/2$ selected from the sum). We note that, with this definition, 
negative-dimensional integrals can be shown to obey the necessary translation
properties 
\cite{HR}.

\subsection{The general case: different masses}
\label{sec:general}
For the one-loop integrals we are interested in here, we follow the
approach suggested by Suzuki and Schmidt~\cite{SS2loop}--\cite{SSbox} and view
Eqs.~(\ref{eq:form1}) and~(\ref{eq:form2}) as existing in negative
dimensions.  
Making the same series expansion of the exponential as above,
Eq.~(\ref{eq:form1}) becomes
\begin{eqnarray}
\Id &=& \Dx
\sum_{n_1,\ldots,n_n=0}^{\infty}
\int 
\frac{d^Dk}{i\pi^{D/2}}
\prod_{i=1}^{n} \frac{(x_iA_i)^{n_i}}{n_i!}\nonumber \\
&=& 
\Dx
\sum_{n_1,\ldots,n_n=0}^{\infty}
I_n^D\(-n_1,\ldots,-n_n;\{Q_i^2\},\{M_i^2\}\)
\prod_{i=1}^n\frac{ x_i^{n_i}}{n_i!},\phantom{aaaa}
\label{eq:LHS}
\end{eqnarray}
where the $n_i$ are positive integers.  The target loop integral is an
infinite sum of (integrals over the Schwinger parameters of) 
loop integrals with negative powers of the propagators.

Likewise, we expand the exponentials in Eq.~(\ref{eq:form2})
\begin{equation}
\label{eq:lhs_expans}
\Id =\Dx
\sum_{n=0}^{\infty}
\frac{\Q^n \P^{-n-\halfD}}{n!} 
\sum_{m=0}^{\infty}
\frac{(-\M)^m}{m!}, 
\end{equation}
and introduce the integers $q_1, \ldots, q_q$, $p_1, \ldots, p_n$
and $m_1,\ldots,m_n$  
to make 
multinomial expansions of $\Q$, $\P$ and $\M$ respectively
\begin{eqnarray}
\label{eq:mult_expans}
\Q^n &=& \sum_{q_{1},\ldots,q_{q}=0}^{\infty}  \frac{\Q_1^{q_1}}{q_1!}
\ldots \frac{\Q_q^{q_q}}{q_q!}\, (q_1+\ldots+q_q)! \nonumber \\
\P^{-n-\halfD} &=& \sum_{p_1,\ldots,p_n=0}^{\infty}  \frac{x_1^{p_1}}{p_1!}
\ldots \frac{x_n^{p_n}}{p_n!} \, (p_1+\ldots+p_n)!  \\
(-\M)^m &=& \sum_{m_1,\ldots,m_n=0}^{\infty}  \frac{\(-x_1M_1^2\)^{m_1}}{m_1!}
\ldots \frac{\(-x_nM_n^2\)^{m_n}}{m_n!} \, (m_1+\ldots+m_n)!,\nonumber
\end{eqnarray}
subject to the constraints 
\begin{equation}
\label{eq:constraints}
\sum_{i=1}^q q_i = n, \quad \quad
\sum_{i=1}^n p_i = -n-\halfD \quad \quad  {\rm and} \quad\quad 
\sum_{i=1}^n m_i = m.
\end{equation}
Altogether, Eqs.~(\ref{eq:lhs_expans}) and~(\ref{eq:mult_expans}) give
\begin{eqnarray}
\label{eq:RHS}
\lefteqn{\Id =} \nonumber \\
&&\Dx
\sum_{{p_1,\ldots,p_n =0 \atop {q_1,\ldots,q_q =0 \atop
m_1,\ldots,m_n=0 }}}^{\infty}
\frac{\Q_1^{q_1}\ldots \Q_q^{q_q}}{q_1!\ldots q_q!}
\frac{x_1^{p_1}\ldots
x_n^{p_n}}{ p_1!\ldots p_n!}
\frac{\(-x_1M_1^2\)^{m_1}}{m_1!}
\ldots \frac{\(-x_nM_n^2\)^{m_n}}{m_n!}
\, (p_1+\ldots+p_n)!, \nonumber \\
\end{eqnarray}
with the constraints expressed by Eq.~(\ref{eq:constraints}).

We recall that each of the $\Q_i$ is a bilinear in the Schwinger parameters, so
that the target loop integral is now an infinite sum of powers of the 
scales of the process (with each of the $M_i^2$ and the $Q_i^2$ raised to a
different summation variable) integrated over the Schwinger parameters.

Equations~(\ref{eq:LHS}) and~(\ref{eq:RHS}) are two different expressions
for the same quantity: $I_n^D$ . 
However, rather than performing the integrals over the $x_i$'s, 
we use the fact that the $x_i$'s are independent parameters, so that the
integrands themselves must be equivalent:
\beqn
\label{eq:lhs_rhs}
&& \hspace{-1cm}\sum_{n_1,\ldots,n_n=0}^{\infty}
I_n^D\(-n_1,\ldots,-n_n;\{Q_i^2\},\{M_i^2\}\)
\prod_{i=1}^n\frac{ x_i^{n_i}}{n_i!} = 
\nonumber\\
&& \sum_{{p_1,\ldots,p_n =0 \atop {q_1,\ldots,q_q =0 \atop
m_1,\ldots,m_n=0 }}}^{\infty}
\frac{\Q_1^{q_1}\ldots \Q_q^{q_q}}{q_1!\ldots q_q!}
\frac{x_1^{p_1}\ldots
x_n^{p_n}}{ p_1!\ldots p_n!}
\frac{\(-x_1M_1^2\)^{m_1}}{m_1!}
\ldots \frac{\(-x_nM_n^2\)^{m_n}}{m_n!}
\, (p_1+\ldots+p_n)!  \phantom{aaaa}
\eeqn
Comparing term by term the left-hand-side (lhs) with the right-hand-side
(rhs) allows us to read off the value of $I_n^D$. In fact, the
coefficient of the term $x_1^{-\nu_1}\ldots x_n^{-\nu_n}$ in the lhs of
Eq.~(\ref{eq:lhs_rhs}), where 
the $\nu_i$ are negative integers, is given by
\begin{equation}
\label{eq:left}
I_2^D(\nu_1,\ldots,\nu_n) \(\prod_{i=1}^n \frac{1}{\Gamma(1-\nu_i)}\).
\end{equation}
This term is equal to the coefficient of the term $x_1^{-\nu_1}\ldots
x_n^{-\nu_n}$ in the rhs of Eq.~(\ref{eq:lhs_rhs}).
Writing a general expression is not possible, since the $\Q_i$ are process
dependent. Nevertheless, we can  extract the
momentum scale $Q_i^2$ from each of the $\Q_i$ and find the coefficient of
this term to be
\beqn
\label{eq:right}
\sum_{{p_1,\ldots,p_n =0 \atop {q_1,\ldots,q_q =0 \atop
m_1,\ldots,m_n=0 }}}^{\infty}
\(Q_1^2\)^{q_1}\ldots\(Q_q^2\)^{q_q}
\(-M_1^2\)^{m_1}\ldots \(-M_n^2\)^{m_n}\phantom{aaaaaa}&& 
\nonumber\\
&& \hspace{-8cm}\times\left(
\prod_{i=1}^{n} \frac{1}{\Gamma(1+m_i)\Gamma(1+p_i)}
\right)
\left(
\prod_{i=1}^{q} \frac{1}{\Gamma(1+q_i)}
\right)
\Gamma\left(1+\sum_{k=1}^n p_k\right), 
\eeqn
subject to the $n$ (process-dependent) constraints that 
ensure that the power of $x_i$ on the lhs ($-\nu_i$) is equal to the
power of $x_i$ on the rhs, which is generally a combination of integers of
the summation. 
 
By adding together the first two expressions in Eq.~(\ref{eq:constraints}),
we obtain an additional constraint, that is
\begin{equation}
\label{eq:power}
p_1+\ldots + p_n +q_1 +\ldots + q_q = -\frac{D}{2}.
\end{equation}

Equating Eqs.~(\ref{eq:left}) and (\ref{eq:right}),
we obtain an expression for the 
loop integral with negative powers of the propagators in negative
dimensions
\begin{eqnarray}
\label{eq:sum}
\Id   
&\equiv&
\sum_{{p_1,\ldots,p_n =0 \atop {q_1,\ldots,q_q =0 \atop
m_1,\ldots,m_n=0 }}}^{\infty}
\(Q_1^2\)^{q_1}\ldots\(Q_q^2\)^{q_q}\,
\(-M_1^2\)^{m_1}\ldots \(-M_n^2\)^{m_n}\nonumber \\
&\times&\left(
\prod_{i=1}^{n} \frac{\Gamma(1-\nu_i)}{\Gamma(1+m_i)\Gamma(1+p_i)}
\right)
\left(
\prod_{i=1}^{q} \frac{1}{\Gamma(1+q_i)}
\right)
\Gamma\left(1+\sum_{k=1}^n p_k\right). \nonumber \\
\end{eqnarray}
Equation~(\ref{eq:sum}), together with the constraints, is the main result of
this paper. The loop integral is written directly as an infinite sum.  Given
that $\Q$ can be read off directly from the Feynman graph, so can the precise
form of Eq.~(\ref{eq:sum}) as well as the system of constraints. Of course,
strictly speaking we have assumed that both $\nu_i$ and $D/2$ are negative
integers and we must be careful in interpreting this result in the physically
interesting domain where the $\nu_i$ and $D$ are all positive. However, this is
relatively straightforward and in the following sections we show how quite
general results for one-loop massive bubbles  and triangles can be obtained.

\begin{itemize}
\item[] {\bf Example:}
to give an explicit example of how Eq.~(\ref{eq:sum}) and the system of
constraints appear, we consider the one-loop bubble with different masses.
Equations~(\ref{eq:LHS}) and~(\ref{eq:RHS}) become
\begin{eqnarray}
\lefteqn{I_2^D(\nu_1,\nu_2;Q_1^2,M_1^2,M_2^2)}\nonumber \\
&=& \Dx
\sum_{n_1,n_2=0}^{\infty}
I_2^D\(-n_1,-n_2;Q_1^2,M_1^2,M_2^2\)
\frac{ x_1^{n_1}x_2^{n_2}}{n_1! \, n_2!}\nonumber \\
&=&\Dx
\sum_{p_1,p_2,q_1,m_1,m_2=0}^{\infty}
\frac{\(x_1x_2\,Q_1^2\)^{q_1} \,
x_1^{p_1}x_2^{p_2} \(-x_1M_1^2\)^{m_1}\(-x_2M_2^2\)^{m_2}}{ 
q_1!\, p_1! \, p_2! \, m_1! \, m_2!} \,
(p_1+p_2)!,  \nonumber\\
\end{eqnarray}
so that, by selecting powers of $x_1^{-\nu_1}$ and $x_2^{-\nu_2}$, we find
(see Eq.~(\ref{eq:sum}))
\begin{eqnarray}
\label{eq:bub1}
I_2^D(\nu_1,\nu_2;Q_1^2,M_1^2,M_2^2)
&=&
\sum_{p_1,p_2,q_1,m_1,m_2=0}^{\infty}  
\(Q_1^2\)^{q_1} \(-M_1^2\)^{m_1} \(-M_2^2\)^{m_2} \nonumber \\
&\times& \frac{\Gamma(1-\nu_1)\Gamma(1-\nu_2)\Gamma(1+p_1+p_2)}
 {\Gamma(1+m_1)\Gamma(1+m_2)\Gamma(1+p_1)\Gamma(1+p_2)\Gamma(1+q_1)},
\phantom{aaaa}
\end{eqnarray}
together with the system of constraints
\begin{eqnarray}
\label{eq:sysbubm1m2}
q_1+p_1 +m_1 &=& -\nu_1,\nonumber \\
q_1+p_2 +m_2 &=& -\nu_2, \\
q_1+p_1+p_2 &=& -\halfD.\nonumber
\end{eqnarray}
In Sec.~\ref{sec:general_form}, we will show how this particular system can
be solved to give results for the bubble integral in positive dimensions $D$,
with arbitrary positive powers of the propagators.
\end{itemize}

\subsection{The general form of the solutions}

In general, for an $n$-point one-loop integral with $q$ external momentum
scales and $m$ mass scales, there will be $(n+q+m)$ summation variables and
$(n+1)$ constraints. Altogether we expect $(n+q+m)! / (n+1)! / (q+m-1)!$
possible solutions (some of which will be eliminated by the specific form of
the system of constraints).  It is easy to see that these solutions span physically
different kinematic regions (depending on the powers of the kinematic scales)
and the summations will only converge in the appropriate kinematic domain.
We expect that solutions in one kinematic region should be
analytically linked to those in other domains.

Each solution of the system of constraints, once inserted  into the template 
of Eq.~(\ref{eq:sum}), has the following generic form
\begin{equation}
\label{eq:gen_form}
\Pre \times \Sum,
\end{equation}
where we have introduced the following notation:
\begin{itemize}
\item[-] 
$\Sum$ is the sum over the terms that contain unconstrained indices of
summation. 
Instead of dealing with $\Gamma$ functions, we have formed \poch\ symbols,
defined as 
\begin{equation}
\label{eq:poch}
(z,n) \equiv \frac{\Gamma(z+n)}{\Gamma(z)},
\end{equation}
because they are the most suitable way to write generalized hypergeometric
functions.
For example, in the case where there is only one remaining summation variable
$n$, then $\Sum$ takes the form 
\begin{equation}
\label{eq:generic}
\Sum \sim \sum_{n=0}^{\infty} \frac{(a_1,n)\ldots (a_N,n)}
{(b_1,n)\ldots(b_{N-1},n)} \, \frac{x^n}{n!} ,
\end{equation}
where $x$ is the ratio of kinematic scales.
The variables $a_i$ and $b_i$ are linear in the $\nu_i$ and $D$ and do not
depend on the summation variables.   
To put $\Sum$ in this form, it is often convenient to use the identity (see
Eq.~(\ref{eq:gamma_flip}))
\begin{equation}
\label{eq:pochflip}
(z,-n) = (-1)^n \frac{1}{(1-z,n)}.
\end{equation} 

In most cases, $\Sum$ can be directly identified as a generalized
hypergeometric function, in the region of convergence of the series.  In
general, these hypergeometric functions are analytic and may be evaluated at
positive values of $D$ and $\nu_i$.

\item[-]
The prefactor $\Pre$ contains all the rest of the terms that are not included
in $\Sum$. More precisely, it is a product of external scales raised to
fixed powers, and $\Gamma$ functions that do not depend on the summation
variables.  These may be produced either directly from the particular
solution of the system, or in the generation of the \poch\ symbols.

In the general case of an $n$-point one-loop integral with $q$ external
momentum scales and $m$ mass scales, inspection of Eq.~(\ref{eq:sum})
dictates that we produce:
\begin{itemize}
\item[-]
$n$ $\Gamma$ functions with argument $(1-\nu_i)$,
\item[-]
$(n+m+q)$ factorials of the summation variables in the denominator,
\item[-]
one $\Gamma$ function in the numerator: $\G{1+\sum_{k=1}^n p_k}$. 
\end{itemize}
Applying the $(n+1)$ constraints leaves $(m+q-1)$ factorials of the remaining
unconstrained summation variables and produces an additional $(n+1)$ $\Gamma$
functions in the denominator and one in the numerator, as \poch\ symbols are
formed using Eq.~(\ref{eq:poch}). Altogether there will be $(n+2)$
\poch\ symbols in $\Sum$, while $\Pre$ will be a ratio with $(n+1)$ $\Gamma$
functions in {\em both} numerator and denominator.  
In both $\Sum$ and $\Pre$, the number of functions may be reduced if there are
cancellations between numerator and denominator.
\end{itemize}
For physical loop integrals with positive powers of propagators, we need to
evaluate $\Pre$ at positive values of the $\nu_i$ and positive $D$. A problem
is immediately obvious: the numerator of $\Pre$ contains $\Gamma(1-\nu_i)$,
so that, for positive integer values $\nu_i$, it appears that we need to
evaluate the $\Gamma$ functions for negative arguments, where they are
singular.  However, $\Pre$ is an analytic function and these singularities
cancel between the numerator and denominator.

In fact, it can be easily shown that, starting from the identity
\begin{equation}
\G{z+1} = z \,\G{z},
\end{equation}
we have
\begin{equation}
\label{eq:gamma_flip}
\frac{\G{z}}{\G{z-n}} = (-1)^{-n} \, \frac{\G{n+1-z}}{\G{1-z}},
\end{equation}
where $z$ is a real (or complex) number,  and $n$ is a positive integer.

In the product of $\Gamma$ functions in the numerator and denominator of the
$\Pre$ term, we can make an iterated use of the
identity~(\ref{eq:gamma_flip}), provided we treat $D/2$ as an integer, as we
have already done in the multinomial expansion.
We can then rewrite the $\Gamma$-function prefactor in a more
amenable way by flipping all of the $\Gamma$ functions from numerator to
denominator and vice versa
\begin{equation}
\prod_{i=1}^{n+1}\frac{\Gamma(\alpha_i)}
     {\Gamma(\beta_i)}  = (-1)^{\sum_{i=1}^{n+1} (\beta_i- \alpha_i)}  
\prod_{i=1}^{n+1}\frac{\Gamma(1-\beta_i)}
{\Gamma(1-\alpha_i)},
\label{eq:flip} 
\end{equation} 
where the index $i$ runs over all $(n+1)$ $\Gamma$ functions in the numerator
and denominator of $\Pre$.
In addition, it can be shown that
\begin{equation} \sum_{i=1}^{n+1}
(\beta_i- \alpha_i) = \frac{D}{2}, 
\label{eq:args}
\end{equation} 
which is independent of the $\nu_i$.

\subsubsection{An example: the massless bubble}

Returning to the example of the one-loop self-energy diagram introduced in
Sec.~\ref{sec:general}, and setting the masses of the internal lines to zero,
$M_1 = M_2 = 0$ (which is equivalent to terminating the series in $m_1$ and
$m_2$ at the first term), we obtain the simpler system of constraints (see
Eq.~(\ref{eq:sysbubm1m2}) with $m_1=m_2=0$)
\begin{eqnarray}
q_1+p_1 &=& -\nu_1,\nonumber \\
q_1+p_2 &=& -\nu_2,\\
q_1+p_1+p_2 &=& -\halfD.\nonumber 
\label{eq:sysbub00}
\end{eqnarray}
Since $m=0$, $q=1$ and $n=2$, we expect that the $(n+1)=3$ constraints 
exactly determine the $(m+q+n)=3$ variables.  In this case $\Sum = 1$ and
the result is entirely given by the prefactor $\Pre$.
Solving this system yields
\begin{eqnarray}
q_1 &=& \halfD-\nu_1-\nu_2,\nonumber \\
p_1 &=& \nu_2-\halfD,\nonumber \\
p_2 &=& \nu_1-\halfD.\nonumber
\end{eqnarray}
Inserting these values directly into Eq.~(\ref{eq:bub1}) with $m_1=m_2=0$
we find
\begin{eqnarray}
I_2^D(\nu_1,\nu_2;Q_1^2,0,0)\!\!\! &=&\! \! \! \Pre \,\ \mbox{}\nonumber\\
&=&\! \! \! 
\frac{\G{1-\nu_1}\G{1-\nu_2}\G{1+\nu_1+\nu_2-D}}
{\G{1+\nu_1-\halfD}
\G{1+\nu_2-\halfD}\G{1+\halfD-\nu_1-\nu_2}}\(Q_1^2\)^{\halfD-\nu_1-\nu_2}.
\nonumber\\
\label{eq:bub2}
\end{eqnarray}
As expected there are $(n+1)$ $\Gamma$ functions in both numerator and
denominator and furthermore the arguments satisfy Eq.~(\ref{eq:args}).
We therefore apply Eq.~(\ref{eq:flip}) and find
\begin{equation}
I_2^D(\nu_1,\nu_2;Q_1^2,0,0) = (-1)^{\halfD}
\frac{\G{\halfD-\nu_1}\G{\halfD-\nu_2}\G{\nu_1+\nu_2-\halfD}}
{\G{ \nu_1}\G{ \nu_2}\G{D-\nu_1-\nu_2}}
   \(Q_1^2\)^{\halfD-\nu_1-\nu_2},
\end{equation}
where $\nu_1$ and $\nu_2$ are positive and 
which agrees with the known result straightforwardly 
obtained using Feynman parameters.

\subsection{Massive bubble integrals}
\label{sec:general_form}
We want now to give a detailed description of how to build the solutions
starting from the general form~(\ref{eq:sum}) for the loop integral and from
the system of constraints, and we want to discuss how the solutions of the
system of constraints need to be combined to give a meaningful result.

We will refer to a precise example to make things clearer: the bubble
integral with different masses in the propagators of Eq.~(\ref{eq:bub1}).
Particular cases with $\nu_1 = \nu_2 = 1$ are important in electroweak
renormalization and have been known for some time (see for
example~\cite{HV}).  The more general cases with $\nu_1 \neq \nu_2 \neq 1$
have been studied by Boos and Davydychev~\cite{BD} using the Mellin-Barnes
integral representations.  

As discussed at the end of Sec.~\ref{sec:general}, in the case where the two
masses are non-zero and different, there are $(m+n+q)=5$ summation variables:
two for the propagator masses ($m_1$ and $m_2$), two for the expansion of $\P$
($p_1$ and $p_2$) and one for the external momentum scale ($q_1$).

The $(n+1)=3$ constraints are given in Eq.~(\ref{eq:sysbubm1m2}). There are a
maximum of $5!/3!/2! = 10$ possible solutions, one for each of the ways in
which we can choose three variables among the five, and solve the system with
respect to these triplets. In this case, 
there is no solution if we try to solve the system for
$\{p_1,q_1,m_2\}$ or $\{p_2,q_1, m_1\}$, so that we have only eight solutions.

Each of the eight solutions corresponds to different values of the integer
summation variables and we insert each of them into the general expression 
for the propagator integral, Eq.~(\ref{eq:bub1}). 
For example, solving for $\{p_1, p_2, q_1\}$, yields
\begin{eqnarray}
\label{eq:p1p2q1}
p_1 &=& \nu_2+m_2-\halfD,\nonumber\\
p_2 &=& \nu_1+m_1-\halfD,\\
q_1 &=& \halfD-\nu_1-\nu_2-m_1-m_2, \nonumber 
\end{eqnarray}
and the contribution of this solution to the integral~(\ref{eq:bub1}) is
\begin{eqnarray}
I_2^{\{m_1, m_2\}}
&=&
\sum_{m_1,m_2=0}^{\infty}  
\(Q_1^2\)^{\halfD-\nu_1-\nu_2-m_1-m_2}\, 
\(-M_1^2\)^{m_1} \,\(-M_2^2\)^{m_2} \;
\frac{\G{1-\nu_1}\G{1-\nu_2}}{\G{1+m_1}\G{1+m_2}}
\nonumber \\
&\times&\frac{
\G{1+\nu_1+\nu_2-D +m_1+m_2}}  
 { \G{1+\nu_2-\halfD+m_2}
\G{1+\nu_1-\halfD +m_1}\G{1+ \halfD-\nu_1-\nu_2-m_1-m_2}},
\nonumber\\
\label{eq:bub1_sol}
\end{eqnarray}
where we have labelled the integral with respect to the indices of summation
and we have dropped the functional dependence of $I_2^{D}$, for ease of
notation. 

As discussed in Sec.~\ref{sec:general}, we now form the \poch\ symbols, and
we make use of the Eq.~(\ref{eq:pochflip}) to flip the \poch\ symbol
in the denominator with negative indices of summation, to obtain 
\begin{eqnarray}
I_2^{\{m_1, m_2\}}\!\!\!\!\!
&=& \!\!\!\(Q_1^2\)^{\halfD-\nu_1-\nu_2} 
\frac{\G{1-\nu_1} \G{1-\nu_2}\G{1+\nu_1+\nu_2-D}}{
\G{1+\nu_2-\halfD}
\G{1+\nu_1-\halfD }\G{1+ \halfD-\nu_1-\nu_2}}  
\nonumber \\
&\times& \!\!\!\!\!\! \!\!\!\sum_{m_1,m_2=0}^{\infty} \!\!\!\!\!
 \frac{\(1+\nu_1+\nu_2-D,m_1+m_2\)
 \(\nu_1+\nu_2-\halfD,m_1+m_2\)}  
 { \(1+\nu_2-\halfD,m_2\)
\(1+\nu_1-\halfD,m_1\)} \,
\frac{\({M_1^2}/{Q_1^2}\)^{m_1}}{m_1 !} \,
\frac{\({M_2^2}/{Q_1^2}\)^{m_2}}{m_2 !} ,
\nonumber\\
\end{eqnarray}
so that we can recognize the general form of Eq.~(\ref{eq:gen_form}): the
first line of the rhs is $\Pre$ while the second is $\Sum$.  By flipping the
$\Gamma$ functions in the prefactor term $\Pre$, using Eq.~(\ref{eq:flip}),
we get
\begin{eqnarray}
\label{eq:bubble_sol}
I_2^{\{m_1,m_2\}}\!\!\!\! &=& \!\!
(-1)^{\halfD}\, \(Q_1^2\)^{\halfD-\nu_1-\nu_2} 
\frac{\G{\halfD-\nu_1} \G{\halfD -\nu_2} 
\G{\nu_1+\nu_2 -\halfD}}  
{\G{\nu_1} \G{\nu_2}\G{D-\nu_1-\nu_2}}
\nonumber \\
& \times& \!\!\!
F_4\( 1+\nu_1+\nu_2-D, \nu_1+\nu_2-\halfD, 1+\nu_1-\halfD,  1+\nu_2-\halfD,
\frac{M_1^2}{Q_1^2} , \frac{M_2^2}{Q_1^2} \),\phantom{aaaaa}
\end{eqnarray}
where we have used the definition of Appell's $F_4$ function given in
Eq.~(\ref{eq:f4_def}).

In the same way, we can obtain the other seven solutions:
\begin{eqnarray}
\label{eq:bubble_sols}
I_2^{\{p_1, m_1\}}  &=& (-1)^{\halfD} {\(Q_1^2\)}^{-\nu_1}\(-M_2^2\)^{\halfD-\nu_2}
\frac{\G{\nu_2-\halfD}}
{\G{\nu_2}} \nonumber \\
&& \times ~~~{F_4}\left(1+\nu_1-\halfD,\nu_1,1+\nu_1-\halfD,1+\halfD-\nu_2,
\frac{M_1^2}{Q_1^2},\frac{M_2^2}{Q_1^2} \right),\nonumber \\
%%%%%
I_2^{\{p_2, m_2 \}} &=&(-1)^{\halfD} {\(Q_1^2\)}^{-\nu_2}\(-M_1^2\)^{\halfD-\nu_1}
\frac{\G{\nu_1-\halfD}}
{\G{\nu_1}} \nonumber \\
&& \times ~~~{F_4}\left(1+\nu_2-\halfD,\nu_2,1+\halfD-\nu_1,1+\nu_2-\halfD,
\frac{M_1^2}{Q_1^2},\frac{M_2^2}{Q_1^2} \right),\nonumber \\
%%%%%
I_2^{\{p_1, p_2 \}} &=& (-1)^{\halfD} 
{\(Q_1^2\)}^{-\halfD}\(-M_1^2\)^{\halfD-\nu_1}\(-M_2^2\)^{\halfD-\nu_2}
\frac{\G{\nu_1-\halfD}\G{\nu_2-\halfD}\G{\halfD}}
{\G{\nu_1}\G{\nu_2} \G{0}} \nonumber \\
&& \times ~~~{F_4}\left(1,\halfD,1+\halfD-\nu_1,1+\halfD-\nu_2,
\frac{M_1^2}{Q_1^2},\frac{M_2^2}{Q_1^2} \right),\nonumber \\
%%%%%
I_2^{\{q_1, m_2 \}} &=& (-1)^{\halfD}\(-M_1^2\)^{\halfD-\nu_1-\nu_2}
\frac{\G{\nu_1+\nu_2-\halfD}\G{\halfD-\nu_2}}
{\G{\nu_1}\G{\halfD}} \nonumber \\
&& \times ~~~{F_4}\left(\nu_1+\nu_2-\halfD,\nu_2,\halfD,1+\nu_2-\halfD,
\frac{Q_1^2}{M_1^2},\frac{M_2^2}{M_1^2} \right),\nonumber \\
%%%%%
I_2^{\{p_1, q_1 \}} &=& (-1)^{\halfD}
\(-M_1^2\)^{-\nu_1}\(-M_2^2\)^{\halfD-\nu_2}
\frac{\G{\nu_2-\halfD}}
{\G{\nu_2}} \nonumber \\
&& \times ~~~{F_4}\left(\nu_1,\halfD,\halfD,1+\halfD-\nu_2,
\frac{Q_1^2}{M_1^2},\frac{M_2^2}{M_1^2} \right),\nonumber \\
%%%%%%
I_2^{\{q1, m_1\}} &=&(-1)^{\halfD} \(-M_2^2\)^{\halfD-\nu_1-\nu_2}
\frac{\G{\nu_1+\nu_2-\halfD}\G{\halfD-\nu_1}}
{\G{\nu_2}\G{\halfD}} \nonumber \\
&& \times ~~~{F_4}\left(\nu_1+\nu_2-\halfD,\nu_1,1+\nu_1-\halfD,\halfD,
\frac{M_1^2}{M_2^2},\frac{Q_1^2}{M_2^2} \right),\nonumber \\
%%%%%
I_2^{\{p_2, q_1 \}} &=& (-1)^{\halfD}
\(-M_1^2\)^{\halfD-\nu_1}\(-M_2^2\)^{-\nu_2}
\frac{\G{\nu_1-\halfD}}
{\G{\nu_1}} \nonumber \\
&& \times ~~~{F_4}\left(\nu_2,\halfD,1+\halfD-\nu_1,\halfD,
\frac{M_1^2}{M_2^2},\frac{Q_1^2}{M_2^2} \right).
\end{eqnarray}

Solution $I_2^{\{p_1, p_2 \}}$ deserves a comment. 
Before any flipping of the $\Gamma$ functions between numerator and
denominator,  the prefactor $\Pre$, has a $\G{1}$ in the numerator, 
due to the fact that the
Appell's $F_4$ function has its first argument equal to 1.  Flipping this
$\G{1}$ according to Eq.~(\ref{eq:flip}) generates a $\G{0}$ in the
denominator, so that this solution is to be considered to be equal to 0.

\subsubsection{Identification of the groups of solutions using the
convergence regions}

The Appell's $F_4(\al,\bt,\ga,\gp,x,y)$, defined in Eq.~(\ref{eq:f4_def}), is
convergent only if (see Table~\ref{tab:convergence})
\begin{equation}
   |\sqrt{x}| + |\sqrt{y}| < 1 ,
\end{equation}
so that we can form three different groups, according to the kinematic
region of convergence of the series
\begin{equation}
\label{eq:i2d_groups}
\begin{array}{lll}
{\displaystyle I_2^D\(\nu_1,\nu_2;Q_1^2,M_1^2,M_2^2\) }&
= {\displaystyle
I_2^{\{m_1,m_2\}}+I_2^{\{p_2,m_2\}}+I_2^{\{p_1,m_1\}} }
&\quad {\rm if} \quad 
 {\displaystyle \sqrt{M_1^2}+\sqrt{M_2^2} < \sqrt{Q_1^2}},\vspace{2mm}
\\ 
{\displaystyle I_2^D\(\nu_1,\nu_2;Q_1^2,M_1^2,M_2^2\) }&= 
 {\displaystyle I_2^{\{q_1,m_2\}}+I_2^{\{p_1,q_1\}} }
&\quad{\rm if} \quad  
 {\displaystyle \sqrt{Q_1^2}+ \sqrt{ M_2^2} < \sqrt{M_1^2}},  \vspace{2mm}
\\ 
{ \displaystyle I_2^D\(\nu_1,\nu_2;Q_1^2,M_1^2,M_2^2\)} &= 
 {\displaystyle I_2^{\{q_1,m_1\}}+I_2^{\{p_2,q_1\}} }
&\quad {\rm if} \quad 
{\displaystyle \sqrt{Q_1^2} + \sqrt{M_1^2} < \sqrt{M_2^2}}.  \vspace{2mm}
\end{array}
\end{equation}
We note that from the convergence properties of the $F_4$, if it was not
eliminated by the zero in the prefactor,  $I_2^{\{p_1,p_2\}}$
would belong to the first kinematic region, $\sqrt{M_1^2}+\sqrt{M_2^2} <
\sqrt{Q_1^2}$.

In this way, using NDIM we have simultaneously obtained {\em all} the different
forms of the hypergeometric functions that express the integral $I_2^D$ for
different kinematic regions of $M_1^2$, $M_2^2$ and $Q_1^2$.  These results for
$I_2^D$ agree with those obtained with the Mellin-Barnes
method~\cite{BD}.

It must be noted that we can go from one kinematic region to the other, just
by applying formula~(\ref{eq:f4_anal_cont}) of the analytic continuation of
the $F_4$ function.
As stated at the end of the previous section, the appearance of $\G{0}$ in
the denominator, that occurs during the process of analytic continuation, 
just kills that term.

\subsubsection{Identification of the groups of solutions using the system of
constraints} 
\label{sec:sys_constr}

We would like to address here a different method to form the groups of
solutions.
This is based only on considerations of the system of constraints, and more
precisely  on the sign of the summed indices of the series $\{p_i, q_i, m_i\}$,
without any knowledge of the region of convergence of the specific series.

For this purpose, the actual value of $\nu_i$ and of $D$ in the
system~(\ref{eq:sysbubm1m2}) is irrelevant, because it only modifies the sign
of a finite number of the summation variables of the series.  For example,
solution~(\ref{eq:p1p2q1}), obtained solving the system with respect of the
two indices $\{m_1, m_2\}$, tells us that the ``bulk'' sign of $p_1$ is equal
to that of $m_2$, because for $m_2$ sufficiently large, the contribution of
$\nu_2-\halfD$ is no longer important.  The same thing happens for $p_2$ and
$q_1$, whose ``bulk'' sign is equal to that of $m_1$ and $-m_1-m_2$,
respectively.

For this reason, instead of considering the full inhomogeneous system, 
we consider the homogeneous one, obtained by setting $\nu_i=0$ and $D=0$.
The system~(\ref{eq:sysbubm1m2}) for the massive bubble then becomes
\beqn
 \label{eq:syshomog}
q_1+p_1 +m_1 &=& 0     ,\nonumber \\
q_1+p_2 +m_2 &=& 0   , \\
q_1+p_1+p_2 &=& 0. \nonumber 
\eeqn
We would like to stress the fact that the last equation has always the same
form, since this is the constraint expressed by Eq.~(\ref{eq:power}).

We can now build a table of signs for $p_i$, $q_i$ and $m_i$.
The last equation gives rise to one of the following cases:
\begin{center}
\leavevmode
%\footnotesize
\begin{tabular}{c| c}
 \hline
$>0$  &  $<0$ \\
\hline
$p_1, \; p_2$ &  $q_1$ \\
$p_1, \; q_1$ &  $p_2$ \\
$p_2, \; q_1$ &  $p_1$ \\
$p_1$ & $p_2, \; q_1$ \\
$p_2$ & $p_1, \; q_1$ \\
$q_1$ & $p_1, \; q_2$ \\
      \hline
\end{tabular}
%\caption{}
\end{center}

Using the last equation of~(\ref{eq:syshomog}) to eliminate $q_1$ from
the other two equations of the system~(\ref{eq:syshomog}), we have
\begin{eqnarray}
\label{eq:m1eqp2}
m_1 &=& p_2     , \nonumber\\ 
m_2 &=& p_1      , 
\end{eqnarray}
so that we can complete the previous table in the following way:
\begin{center}
\begin{tabular}{c| c}
 \hline
$>0$  &  $<0$ \\
 \hline
$p_1, \; p_2, \; m_1, \; m_2$ &  $q_1$ \\
$p_1, \; q_1, \; m_2$ &  $p_2, \; m_1$ \\
$p_2, \; q_1, \; m_1$ &  $p_1, \; m_2$ \\
      \hline
\end{tabular}
\end{center}
\noindent where the last three lines have been neglected, being
equal and opposite to the first three ones.

We know that the summation indices of the series must be positive integers and
we can therefore imagine solving the system~(\ref{eq:sysbubm1m2}) with respect
to any pair of variables that are simultaneously positive.
The table provides us with this information and  
we can directly read from the table 
which integers are simultaneously positive and use them to 
form a group of solutions with similar properties 
by selecting all possible pairs of summation indices from the list of positive
indices.
 
Starting from the first row of the table, and considering the
identities~(\ref{eq:m1eqp2}), that embody the fact that we cannot solve the
system with respect to the pairs of indices $\{p_1, m_2\}$ and $\{p_2,
m_1\}$, because they are linearly dependent, we can form the following
subgroups:
\begin{equation}
p_1, \; p_2, \; m_1, \; m_2  \quad \Longrightarrow \quad
\{p_1, p_2\}, \;
\{p_1, m_1\}, \;
\{p_2, m_2\}, \;
\{m_1, m_2\}.
\end{equation}
The same thing can be done with the other two rows of the table:
\beqn
p_1, \; q_1, \; m_2 \quad &\Longrightarrow& \quad
\{p_1, q_1\}, \;
\{q_1, m_2\}, \\
p_2, \; q_1, \; m_1  \quad &\Longrightarrow& \quad
\{p_2, q_1\}, \;
\{q_1, m_1\}. 
\eeqn
These are exactly the groups obtained by adding solutions according to their
region of convergence (see Eq.~(\ref{eq:i2d_groups})), once we consider the
fact that $I_2^{\{p_1, p_2\}}=0$.

This method gives the correct groups only for the cases where the homogeneous
system can be solved without any ambiguity.
There are examples, and we will meet one in Sec.~\ref{sec:m3=0}, where the
sign of some indices of the series are undetermined, because they have a
dependence on other indices of the type
 \[
    p_1 = p_2 + p_3,
 \]
where $p_2$ and $p_3$ have opposite sign. In this case, we cannot say if 
$p_1$ is positive or negative.

Up to now, we do not have a way to deal with these cases directly from the
system of constraints, and we leave the task of further investigating this
issue to future works.

\subsubsection[The limiting case: $M_1 \neq 0,\, M_2 = 0$]
{The limiting case: $\boldsymbol{M_1 \neq 0,\, M_2 = 0}$}
We conclude this section, by considering some extreme cases.
First we consider the limit of one massless
propagator in the self-energy diagram.
We can compute this integral in two different ways.
\begin{itemize}
  \item[1.]  We can start with the system~(\ref{eq:sysbubm1m2}) with $m_2=0$:
   we have four variables and three constraints, so that we end
   up with a single-index series, that turns out to be a Gauss' hypergeometric
   ${_2F_1}$ function (see Eq.~(\ref{eq:f21_def})).
  
  \item[2.] We can simply take the limit for $M_2 \,\to\, 0$ of the general
   expressions~(\ref{eq:bubble_sol}) and~(\ref{eq:bubble_sols}).  We can
   apply this limit only to the solutions that are convergent in the new
   kinematic regions, and we cannot take the limit for solutions
   $I_2^{\{q_1,m_1\}}$ and $I_2^{\{p_2,q_1\}}$ because they are defined only
   for $\sqrt{Q_1^2} + \sqrt{M_1^2} < \sqrt{M_2^2}$.
   
   The expression for $F_4(\al,\bt,\ga,\gp,x,0)$ is easily obtained from its
   definition~(\ref{eq:f4_def}) with the second summation series collapsing to
   its first term
   \begin{equation} 
      F_4\(\al,\bt,\ga,\gp,x,0\) = \f21\(\al,\bt,\ga,x\).
   \end{equation}
\end{itemize}
Both procedures give the same result.
\begin{eqnarray}
\label{eq:sol_bub_q1}
{\rm If\ }  M_1^2 < Q_1^2 \hspace{2cm}  && \nonumber\\
I_2^D\(\nu_1,\nu_2;Q_1^2,M_1^2,0\) &=& I_2^{\{m_1\}} +
I_2^{\{p_2\}} \,  
\nonumber \\
&=& (-1)^{\halfD}\left(Q_1^2\right)^{\halfD-\nu_1-\nu_2}
\frac{\G{\nu_1+\nu_2-\halfD}\G{\halfD-\nu_1}\G{\halfD-\nu_2}}
{\G{\nu_1}\G{\nu_2}\G{D-\nu_1-\nu_2}} \nonumber \\
&\times &{_2F_1}\left(1+\nu_1+\nu_2-D,\nu_1+\nu_2-\halfD,
1+\nu_1-\halfD,
\frac{M_1^2}{Q_1^2} \right)\nonumber \\
&+& (-1)^{\halfD} \left(Q_1^2\right)^{-\nu_2}\(-M_1^2\)^{\halfD-\nu_1}
\frac{\G{\nu_1-\halfD}}
{\G{\nu_1}} \nonumber \\
& \times & {_2F_1}\left(\nu_2,1+\nu_2-\halfD,1+\halfD-\nu_1,
\frac{M_1^2}{Q_1^2} \right) , 
\\
{\rm if\ } Q_1^2 < M_1^2  \hspace{2cm}   \nonumber\\
\label{eq:sol_bub_m1}
I_2^D\(\nu_1,\nu_2;Q_1^2,M_1^2,0\) &=& I_2^{\{q_1\}} \,  \nonumber \\
&=& (-1)^{\halfD} \(-M_1^2\)^{\halfD-\nu_1-\nu_2}
\frac{\G{\nu_1+\nu_2-\halfD}\G{\halfD-\nu_2}}
{\G{\nu_1}\G{\halfD}} 
\nonumber\\
&\times& {_2F_1}\left(\nu_1+\nu_2-\halfD,\nu_2,\halfD,
\frac{Q_1^2}{M_1^2} \right).
\end{eqnarray}

\subsubsection[The limiting case: $Q_1^2 \,\to\, 0$]
{The limiting case: $\boldsymbol{Q_1^2 \,\to\, 0}$}

Similarly, we can take the limit of the general self-energy diagram where the
external momentum scale vanishes.  Once again, we can either return to the
system~(\ref{eq:sysbubm1m2}) with one fewer variable ($q_1=0$) or we just
take the $Q_1^2 \,\to\, 0$ limit of the general result~(\ref{eq:i2d_groups}) in
the appropriate kinematic regions: $\sqrt{Q_1^2} + \sqrt{M_1^2} <
\sqrt{M_2^2}$ or $\sqrt{Q_1^2} + \sqrt{M_2^2} < \sqrt{M_1^2}$.  Both
procedures yield the same result:
\begin{eqnarray}
\label{eq:sol_bub_q1to0}
\lefteqn{{\rm If\ } M_1 > M_2\hspace{1.5cm}} \nonumber\\
\lefteqn{I_2^D(\nu_1,\nu_2;0,M_1^2,M_2^2)=
I_2^{\{ m_2\}}+I_3^{\{ p_1\}} 
}\nonumber \\
&=& (-1)^{\halfD}\(-M_1^2\)^{\halfD-\nu_1-\nu_2}
\frac{\G{\nu_1+\nu_2-\halfD}\G{\halfD-\nu_2}}
{\G{\nu_1}\G{\halfD}} \nonumber \\
&& \times ~{_2F_1}\left(\nu_2,\nu_1+\nu_2-\halfD,1+\nu_2-\halfD,
\frac{M_2^2}{M_1^2} \right)\nonumber \\
%%%%%
&+& (-1)^{\halfD}
\(-M_1^2\)^{-\nu_1}\(-M_2^2\)^{\halfD-\nu_2}
\frac{\G{\nu_2-\halfD}}
{\G{\nu_2}} ~{_2F_1}\left(\nu_1,\halfD,1+\halfD-\nu_2,
\frac{M_2^2}{M_1^2} \right), \phantom{aaaa}
\end{eqnarray}
with the result for $M_2 > M_1$ obtained by the exchanges 
$M_1 \leftrightarrow M_2$ and $\nu_1 \leftrightarrow \nu_2$.

Provided that we do not violate the validity of the kinematic regions, we can
take the subsequent limits of the energy scales.  For example, we can safely
take the $M_2 \,\to\, 0$ limit for the solution where $M_2 < M_1$. In this
case, only the first term survives and we obtain the familiar result
\begin{equation}
\label{eq:sol_bub_q1m2to0}
I_2^D(\nu_1,\nu_2;0,M_1^2,0)=
(-1)^{\halfD}\(-M_1^2\)^{\halfD-\nu_1-\nu_2}
\frac{\G{\nu_1+\nu_2-\halfD}\G{\halfD-\nu_2}}
{\G{\nu_1}\G{\halfD}}.
\end{equation}

\subsection{The special case: all masses equal}
\label{subsec:equal}

For the special case where each propagator has the same mass,
Eq.~(\ref{eq:M}) becomes
\begin{equation}
\M = \sum_{i=1}^n x_i M^2 = \P M^2,
\end{equation}
and we have  an  important simplification.
As before, we expand the exponentials in Eq.~(\ref{eq:form2})
and make multinomial expansions of $\Q$ and $\P$
\begin{eqnarray}
\label{eq:RHSequal}
\Ideq &=&\Dx
\sum_{n=0}^{\infty} \sum_{m=0}^{\infty}
\frac{\Q^n \P^{-n-\halfD+m} (-M^2)^m}{n!\,m!}  \nonumber\\
&=&\Dx
\sum_{{p_1,\ldots,p_n =0 \atop {q_1,\ldots,q_q =0 \atop
m=0 }}}^{\infty}
\frac{\Q_1^{q_1}\ldots \Q_q^{q_q}}{q_1!\ldots q_q!}
\frac{x_1^{p_1}\ldots
x_n^{p_n}}{ p_1!\ldots p_n!}
\frac{(-M^2)^{m}}{m!}
\, (p_1+\ldots+p_n)!,\nonumber\\
\end{eqnarray}
subject to the constraints 
\begin{equation}
\label{eq:constr_equal_masses}
\sum_{i=1}^q q_i = n \quad \quad {\rm and} \quad\quad 
\sum_{i=1}^n p_i = -n-\halfD +m.\quad \quad 
\end{equation}
Equating Eqs.~(\ref{eq:LHS}) and~(\ref{eq:RHSequal}) and, once again,
identifying powers of $x_i^{-\nu_i}$, we obtain an expression for the loop
integral with negative powers of the propagators in negative dimensions for
all masses equal
\begin{eqnarray}
\Ideq
&=&
\sum_{{p_1,\ldots,p_n =0 \atop {q_1,\ldots,q_q =0 \atop
m=0 }}}^{\infty}
%\sum_{p_1,\ldots,p_n,q_1,\ldots,q_q,m=0}^{\infty}  
\(Q_1^2\)^{q_1}\ldots \(Q_q^2\)^{q_q}
\(-M^2\)^{m}\nonumber \\
&& \mbox{}\times\left(
\prod_{i=1}^{n} \frac{\G{1-\nu_i}}{\G{1+p_i}}
\right)
\left(
\prod_{i=1}^{q} \frac{1}{\G{1+q_i}}
\right)
\frac{\G{1+\sum_{k=1}^n p_k}}{\G{1+m}}, \phantom{aaa}
\label{eq:sumequal}
\end{eqnarray}
subject to $n$ constraints that each of the powers of $x_i$ match up correctly.
However, the constraint that matches up the powers of $\Q$ and $\P$, obtained
by summing the two expressions in Eq.~(\ref{eq:constr_equal_masses}),
is now
\begin{equation}
p_1+\ldots + p_n +q_1 +\ldots + q_q = -\frac{D}{2}+m,
\label{eq:powerequal}
\end{equation}
rather than Eq.~(\ref{eq:power}).

We see that there are $(n+q+1)$ summation variables and $(n+1)$ constraints,
leaving $q$ remaining summations. We note that the structure of the solution
is precisely as for the unequal-mass case and is treated in the same way by
constructing the sum over \poch\ symbols $\Sum$ and the $\Gamma$ function
prefactor $\Pre$.

\begin{itemize}
\item[] {\bf Example:}
to give an explicit example, we consider the self-energy correction to the
propagator integral with equal masses.  There are $(n+q+1) = 4$ summation
variables with $(n+1) = 3$ constraints.  In this case, the template solution
Eq.~(\ref{eq:sumequal}) is given by
\beqn
I_2^D\(\nu_1,\nu_2;Q_1^2,M^2,M^2\) &=& 
\sum_{p_1,p_2,q_1,m=0}^{\infty}  
\(Q_1^2\)^{q_1}
\(-M^2\)^{m} \nonumber\\
&&\mbox{}\times
\frac{\G{1-\nu_1}\G{1-\nu_2}\G{1+p_1+p_2}}
{\G{1+p_1}\G{1+p_2}\G{1+q_1}\G{1+m}}, \phantom{aaa}
\label{eq:bubequal}
\eeqn
while, matching the powers of $x_i$ gives the system of constraints
\begin{eqnarray}
\label{eq:sysbubequal}
q_1+p_1 &=& -\nu_1,\nonumber \\
q_1+p_2 &=& -\nu_2, \\
q_1+p_1+p_2 &=& -\halfD+m.\nonumber
\end{eqnarray}
There are four summation variables ($p_1$, $p_2$, $q_1$ and $m$) and three
constraints, and we obtain four series solutions, with only one index of 
summation.

Defining $\sigma=\nu_1+\nu_2$ (see Eq.~(\ref{eq:sigma})), we have:
\begin{eqnarray}
\label{eq:sol_bub_eqm3}
\lefteqn{{\rm If\ } Q_1^2 > 4 M^2 \hspace{1.5cm}} \nonumber\\
\lefteqn{I_2^D\(\nu_1,\nu_2;Q_1^2,M^2,M^2\) = I_2^{\{m\}}+I_2^{\{p_1\}}+I_2^{\{p_2\}}
\,  } 
\nonumber \\
&=& (-1)^{\halfD}\left(Q_1^2\right)^{\halfD-\nu_1-\nu_2}
\frac{\G{\sigma-\halfD}\G{\halfD-\nu_1}\G{\halfD-\nu_2}}
{\G{\nu_1}\G{\nu_2}\G{D-\nu_1-\nu_2}} \nonumber \\
&& \times \ \ {_3F_2}\left(1+\halfs-\halfD,\frac{1}{2}+\halfs-\halfD,
\sigma-\halfD,1+\nu_1-\halfD,1+\nu_2-\halfD,
\frac{4M^2}{Q_1^2} \right)\nonumber \\
&+& (-1)^{\halfD}\left(Q_1^2\right)^{-\nu_1}\(-M^2\)^{\halfD-\nu_2}
\frac{\G{\nu_2-\halfD}}
{\G{\nu_2}} \nonumber \\
&& \times \ \ {_3F_2}\left(\nu_1,1+\frac{\nu_1}{2}-\frac{\nu_2}{2},
\frac{1}{2}+\frac{\nu_1}{2}-\frac{\nu_2}{2},
1+\nu_1-\nu_2,1+\halfD-\nu_2,
\frac{4M^2}{Q_1^2} \right)\nonumber \\
&+& (-1)^{\halfD} \left(Q_1^2\right)^{-\nu_2}\(-M^2\)^{\halfD-\nu_1}
\frac{\G{\nu_1-\halfD}}
{\G{\nu_1}} \nonumber \\
&& \times \ \ {_3F_2}\left(\nu_2,1+\frac{\nu_2}{2}-\frac{\nu_1}{2}, 
\frac{1}{2}+\frac{\nu_2}{2}-\frac{\nu_1}{2},
1+\nu_2-\nu_1,1+\halfD-\nu_1,
\frac{4M^2}{Q_1^2} \right), \nonumber\\ 
&&\\
\label{eq:sol_bub_eqm1}
\lefteqn{{\rm if\ } Q_1^2 < 4 M^2 \hspace{1.5cm}} \nonumber\\
\lefteqn{I_2^D\(\nu_1,\nu_2;Q_1^2,M^2,M^2\) = I_2^{\{q_1\}} \,} \nonumber\\
&=& (-1)^{\halfD}\(-M^2\)^{\halfD-\nu_1-\nu_2}
\frac{\G{\sigma-\halfD}}
{\G{\sigma}} \ {_3F_2}\left(\nu_1,\nu_2,\sigma-\halfD,
\halfs,\frac{1+\sigma}{2},
\frac{Q_1^2}{4M^2} \right).\nonumber \\
\end{eqnarray}
In forming the \poch\ symbols we have made use of the following duplication
formula
\begin{equation}
\label{eq:poch2n}
  (z,2n) = 4^{n} \(\frac{z}{2},n\) \(\frac{z}{2} + \frac{1}{2}, n\).
\end{equation}

The procedure described in Sec.~\ref{sec:sys_constr} on how the solutions
group together to give the correct answer in a particular kinematic region,
is straightforward.  In fact, the solution of the homogeneous counterpart 
of the system of constraints~(\ref{eq:sysbubequal}) is
\begin{equation}
   p_1=p_2=m=-q_1.
\end{equation}
We can then form groups for the indices of summation of the series out of
combinations of indices with the same sign, that is:
\begin{equation}
  \{p_1\},\;   \{p_2\}, \;  \{m\} \quad \quad {\rm or} \quad\quad \{q_1\}.
\end{equation} 
Not surprisingly these were the groups of 
solutions formed by considering the convergence properties 
in Eqs.~(\ref{eq:sol_bub_eqm3}) and~(\ref{eq:sol_bub_eqm1}) and reproduce the
known result~\cite{BD}..

\end{itemize}

%%%%%%%%%%%%%%%%%%%%%%%%%%%%%%%%%%%%%%%%%%%%%%%%% 
\section{Massive Vertex integrals} 
\label{sec:triangles}
%%%%%%%%%%%%%%%%%%%%%%%%%%%%%%%%%%%%%%%%%%%%%%%%% 

\begin{figure}
\begin{center}
~\epsfig{file=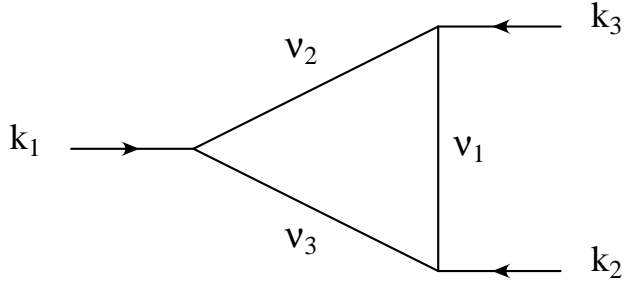,height=4cm}
\end{center}
\caption[]{The one-loop vertex diagram}
\label{fig:tri}
\end{figure}

We now turn to massive-triangle integrals where each propagator
can have a different mass and each external leg can be off-shell.    
The propagators and momenta are labelled as in Fig.~\ref{fig:tri}.
Throughout this section, the number of propagators $n$ is equal to three and
in the most general case, we have
\begin{eqnarray}
\P &=& x_1+x_2+x_3\nonumber \\
\Q &=& x_2x_3\, Q_1^2 + x_3 x_1\, Q_2^2 +x_1 x_2\, Q_3^2\\
\M &=& x_1\, M_1^2 + x_2\, M_2^2 + x_3\, M_3^2, \nonumber
\end{eqnarray}
where $Q_i^2 = k_i^2$.

Based on the discussion of Sec.~\ref{sec:general}, the NDIM method provides 
a generic solution with $(n+q+m)=9$ summation variables and
$(n+1)=4$ constraints, with the template solution given by
\begin{eqnarray}
\lefteqn{I_3^D\(\nu_1,\nu_2,\nu_3;Q_1^2,Q_2^2,Q_3^2,M_1^2,M_2^2,M_3^2\)}
\nonumber \\   
&\equiv&
\sum_{{p_1,\ldots,p_3 =0 \atop {q_1,\ldots,q_3 =0 \atop
m_1,\ldots,m_3=0 }}}^{\infty}
\(Q_1^2\)^{q_1}\(Q_2^2\)^{q_2}\(Q_3^2\)^{q_3}
\(-M_1^2\)^{m_1}\(-M_2^2\)^{m_2} \(-M_3^2\)^{m_3}\nonumber \\
&&\times \left(
\prod_{i=1}^{3} \frac{\Gamma\(1-\nu_i\)}{\Gamma\(1+m_i\)\Gamma\(1+p_i\)
\Gamma\(1+q_i\)}\right)
\Gamma\left(1+p_1+p_2+p_3\right), 
\label{eq:trigen}
\end{eqnarray}
while the system of four constraints is
\begin{eqnarray}
q_2+q_3+p_1 +m_1 &=& -\nu_1,\nonumber \\
q_1+q_3+p_2 +m_2 &=& -\nu_2,\nonumber \\
q_1+q_2+p_3 +m_3 &=& -\nu_3,\nonumber \\
p_1+p_2+p_3+q_1+q_2+q_3 &=& -\halfD.
\label{eq:systrim1m2m3}
\end{eqnarray}
The number of possible solutions satisfying this system is $9!/4!/5! = 126$
of which 45 are eliminated by the particular nature of the system leaving 81.
As usual, insertion of these solutions into the template yields contributions
of the general form (\ref{eq:gen_form}), where $\Sum$ is a product of \poch\
symbols and ratios of energy scales summed over the five remaining variables.
The prefactor $\Pre$ vanishes in a further 12 instances, leaving 69 solutions
which are distributed among the various kinematic regions.

At present, the technology for dealing with five-fold sums (and their
integral representations) is not sufficiently developed to handle the
completely general case.  For the remainder of this section, we therefore
concentrate on particular cases of the vertex integral where some of the
energy scales vanish, leading to either single or double sums which have been
well studied.

\subsection[Massless propagators: ${M_1 =M_2 = M_3= 0}$]
{Massless propagators: $\boldsymbol{M_1 =M_2 = M_3= 0}$}

We first consider the special case where all of the internal lines are massless,
so $m_1 = m_2 = m_3 = 0$ in Eq.~(\ref{eq:systrim1m2m3}),
leaving six summation variables.
Of the $6!/4!/2!=15$ possible solutions of this system, three are eliminated 
by the system,  leaving twelve.
The three-mass triangle is an extremely symmetric system and there are three
allowed phase space regions:
\begin{equation}
\begin{array}{ll}
{\rm region\ I:} & \displaystyle{\sqrt{Q_1^2} > \sqrt{Q_2^2}+\sqrt{Q_3^2},}
\nonumber \vspace{1mm}\\  
{\rm region\ II:} &  \displaystyle{\sqrt{Q_2^2} > \sqrt{Q_1^2}+\sqrt{Q_3^2},} 
\vspace{1mm}\\  
{\rm region\ III:} & \displaystyle{\sqrt{Q_3^2} > \sqrt{Q_1^2}+\sqrt{Q_2^2},}
\\ \nonumber  
\end{array}
\end{equation}
with each region bounded by the external phase-space constraints
\begin{equation}
\Delta_3\(Q_1^2,Q_2^2,Q_3^2\) > 0,
\end{equation}
where
\begin{equation}
\Delta_3(x,y,z) = x^2+y^2+z^2-2xy-2yz-2zx.
\end{equation}
The twelve solutions populate the three
kinematic regions equally, four in each, and are easily identified as belonging
to a particular region by studying either the convergence properties 
of the double sum or by considering the system, 
as in Sec.~\ref{sec:sys_constr}.
For example, the four solutions belonging to region I
($q_1$ negative, $q_2$ and $q_3$ positive) are those where the summation
variables include pairs in the set $\{p_2,p_3,q_2,q_3\}$, that is
$\{p_2,p_3\}$, $\{p_2,q_3\}$,  $\{q_2,p_3\}$ and $\{q_2,q_3\}$, where 
$\{p_2,q_2\}$ and $\{p_3,q_3\}$ have been eliminated by the system.

As usual, each solution is inserted into Eq.~(\ref{eq:trigen}) and treated
according to the procedure described in Sec.~\ref{sec:general}:
the summation variables are converted into \poch\ 
symbols; the $\Gamma$-function prefactor is flipped using
Eq.~(\ref{eq:flip}) and  the remaining summations converted into
generalised hypergeometric functions.  
In each case, we identify Appell's  
$F_4\(\al,\bt,\ga,\gp,x,y\)$ function (see Eq.~\ref{eq:f4_def}),
which, according to the convergence criteria of Table~\ref{tab:convergence},
is well defined when $\sqrt{x}+\sqrt{y} < 1$, precisely matching on to
the physically allowed phase space. 

Summing the four solutions we find, in the region  $\sqrt{Q_1^2} >
\sqrt{Q_2^2}+\sqrt{Q_3^2}$,  
\begin{eqnarray}
\lefteqn{
I_3^D\left(\nu_1,\nu_2,\nu_3;Q_1^2,Q_2^2,Q_3^2,0,0,0\right) = 
I_3^{\{q_2,q_3\}} +
I_3^{\{p_2,q_3\}} +
I_3^{\{p_3,q_2\}} +
I_3^{\{p_2,p_3\}}} \nonumber \\
&=& 
(-1)^{\halfD}
   ~\(Q_1^2\)^{\frac{D}{2}-\nu_1-\nu_2-\nu_3} 
\frac{\Gamma\left(\frac{D}{2}-\nu_1-\nu_2\right)
\Gamma\left(\frac{D}{2}-\nu_1-\nu_3\right)
\Gamma\left(\sigma-\frac{D}{2}\right)}
{\Gamma\left(\nu_2\right)\Gamma\left(\nu_3\right)
\Gamma\left(D-\sigma\right)} \nonumber \\
& &\hspace{1cm} \times  
~~F_4\left(\nu_1,\sigma-\frac{D}{2},1+\nu_1+\nu_3-\frac{D}{2},
 1+\nu_1+\nu_2-\frac{D}{2},\frac{Q_2^2}{Q_1^2},\frac{Q_3^2}{Q_1^2}\right)
\nonumber \\ 
&+&
\left(-1\right)^{\halfD}
   ~\left(Q_1^2\right)^{-\nu_2}
   ~\left(Q_2^2\right)^{\frac{D}{2}-\nu_1-\nu_3}
\frac{\Gamma\left(\frac{D}{2}-\nu_1-\nu_2\right)
\Gamma\left(\nu_1+\nu_3-\frac{D}{2}\right)
\Gamma\left(\frac{D}{2}-\nu_3\right)}
{\Gamma\left(\nu_1\right)
\Gamma\left(\nu_3\right)\Gamma\left(D-\sigma\right)}
\nonumber \\
&& \hspace{1cm} \times  
~~F_4\left(\nu_2,\frac{D}{2}-\nu_3,1+\frac{D}{2}-\nu_1-\nu_3,
1+\nu_1+\nu_2-\frac{D}{2},\frac{Q_2^2}{Q_1^2},\frac{Q_3^2}{Q_1^2}\right) 
\nonumber \\
&+&
\left(-1\right)^{\halfD}
   ~\left(Q_1^2\right)^{-\nu_3}
   ~\left(Q_3^2\right)^{\frac{D}{2}-\nu_1-\nu_2}
\frac{\Gamma\left(\frac{D}{2}-\nu_1-\nu_3\right)
\Gamma\left(\nu_1+\nu_2-\frac{D}{2}\right)
\Gamma\left(\frac{D}{2}-\nu_2\right)}
{\Gamma\left(\nu_1\right)
\Gamma\left(\nu_2\right)\Gamma\left(D-\sigma\right)}
\nonumber \\
&&\hspace{1cm}\times  
~~F_4\left(\nu_3,\frac{D}{2}-\nu_2,1+\nu_1+\nu_3-\frac{D}{2},
1+\frac{D}{2}-\nu_1-\nu_2,\frac{Q_2^2}{Q_1^2},\frac{Q_3^2}{Q_1^2}\right) 
\nonumber \\
&+&
\left(-1\right)^{\halfD}
   ~\left(Q_1^2\right)^{\nu_1-\halfD}
   ~\left(Q_2^2\right)^{\frac{D}{2}-\nu_1-\nu_3}
   ~\left(Q_3^2\right)^{\frac{D}{2}-\nu_1-\nu_2}
   \nonumber \\
   && \hspace{1cm}\times
~~\frac{\Gamma\left(\nu_1+\nu_2-\frac{D}{2}\right)
\Gamma\left(\nu_1+\nu_3-\frac{D}{2}\right)
\Gamma\left(\frac{D}{2}-\nu_1\right)}
{\Gamma\left(\nu_1\right)
\Gamma\left(\nu_2\right)\Gamma\left(\nu_3\right)}
\nonumber \\
&&  \hspace{1cm}\times 
~~F_4\left(D-\sigma,\frac{D}{2}-\nu_1,1+\frac{D}{2}-\nu_1-\nu_3,
1+\frac{D}{2}-\nu_1-\nu_2,\frac{Q_2^2}{Q_1^2},\frac{Q_3^2}{Q_1^2}\right) 
,\phantom{aaaa} 
\label{eq:tri3result}
\end{eqnarray}
which agrees with that obtained 
by Boos and Davydychev \cite{BD} using the Mellin-Barnes integral
representation.
Similar results are obtained for the other two kinematic regions, either by
directly  summing the solutions valid in that region 
(pairs from $\{p_1,p_3,q_1,q_3\}$ or $\{p_1,p_2,q_1,q_2\}$, with 
$\{p_1,q_1\}$, $\{p_2,q_2\}$ and $\{p_3,q_3\}$ excluded by the system)
or by analytic
continuation of the  Appell's $F_4$ function using
formula~(\ref{eq:f4_anal_cont}).  

Note that if one of the $\nu_i$ vanishes (equivalent to propagator $i$
shrinking to a point), only a single term remains.  For example, if $\nu_1 =
0$, only the first term of Eq.~(\ref{eq:tri3result})
survives (the others being killed by $1/\G{0}$), and
the Appell's function collapses to
$$
F_4\(0,\bt,\ga,\gp,x,y\) = 1,
$$
as can be seen from the definition~(\ref{eq:f4_def}), yielding
\begin{equation}
I_3^D\left(0,\nu_2,\nu_3;Q_1^2,Q_2^2,Q_3^2,0,0,0\right) 
= I_2^D\left(\nu_2,\nu_3;Q_1^2,0,0\right), 
\end{equation}
as it should.

We can obtain some other interesting limits if we set to zero one or two
external invariants.

\begin{itemize}
\item[1.] {\bf One light-like external momentum:}
if the $i$th external leg is light-like ($Q_i^2=0$), we can return to the
general case and solve the system~(\ref{eq:systrim1m2m3}) with $q_i=0$, or we
can take the appropriate limit of the general solution.
These limits can be safely made provided that we start from a valid 
kinematic region.
In the region of validity of Eq.~(\ref{eq:tri3result}), that is $\sqrt{Q_1^2}
> \sqrt{Q_2^2}+\sqrt{Q_3^2}$, we can surely take the limits for $Q_2^2 \,\to\, 0$
or $Q_3^2 \,\to\, 0$. In this last case, for example, the last two terms in 
Eq.~(\ref{eq:tri3result}) vanish, while the first two terms collapse to 
Gaussian hypergeometric functions, according to Eq.~(\ref{eq:F4_ytozero}), 
yielding, in the region $Q_1^2 > Q_2^2$, 
\begin{eqnarray}
\lefteqn{
I_3^D(\nu_1,\nu_2,\nu_3;Q_1^2,Q_2^2,0,0,0,0) = 
I_3^{\{q_2\}} +
I_3^{\{p_2\}} } \nonumber \\
&=&(-1)^{\frac{D}{2}}
   ~\(Q_1^2\)^{\frac{D}{2}-\nu_1-\nu_2-\nu_3}
\frac{\Gamma\left(\frac{D}{2}-\nu_1-\nu_2\right)
\Gamma(\frac{D}{2}-\nu_1-\nu_3)\Gamma(\nu_1+\nu_2+\nu_3-\frac{D}{2})}
{\Gamma(\nu_2)\Gamma(\nu_3)\Gamma(D-\nu_1-\nu_2-\nu_3)}
\nonumber \\
&& 
\hspace{3cm}
\times 
\phantom{~}{_2F_1}\left(\nu_1,\sigma-\frac{D}{2},
1+\nu_1+\nu_3-\frac{D}{2},\frac{Q_2^2}{Q_1^2}\right)
      \nonumber \\
&+&(-1)^{\frac{D}{2}}
   ~\(Q_1^2\)^{-\nu_2}
   ~\(Q_2^2\)^{\frac{D}{2}-\nu_1-\nu_3}
\frac{\Gamma\left(\frac{D}{2}-\nu_1-\nu_2\right)
\Gamma\left(\frac{D}{2}-\nu_3\right)
\Gamma\left(\nu_1+\nu_3-\frac{D}{2}\right)}
{\Gamma(\nu_1)\Gamma(\nu_3)\Gamma(D-\nu_1-\nu_2-\nu_3)}
\nonumber \\
&& 
\hspace{3cm} \times
\phantom{~}{_2F_1}\left(\nu_2,\frac{D}{2}-\nu_3,
               1+\frac{D}{2}-\nu_1-\nu_3,\frac{Q_2^2}{Q_1^2}\right)
~. \phantom{aaa}
\label{eq:tri2result} 
\end{eqnarray}

Analogous results valid in the region
$Q_2^2 > Q_1^2$ can be obtained either by starting from
the expression for $I_3^D$ in region II , that is $\sqrt{Q_2^2} >
\sqrt{Q_1^2}+\sqrt{Q_3^2}$,  or via analytic
continuation of Eq.~(\ref{eq:tri2result}), according to
Eq.~(\ref{eq:f21_anal_1/z}).

\item[2.] {\bf Two light-like external momenta:}
in a similar way,  we can obtain the result for two light-like
external momenta, $Q_3^2=Q_2^2 = 0$ for example, by simultaneously taking 
both $Q_2^2$ and $Q_3^2 \,\to\, 0$ in Eq.~(\ref{eq:tri3result}).   
Only the first term in Eq.~(\ref{eq:tri3result})
survives, and the Appell's function collapses to
\begin{equation}
F_4\(\al,\bt,\ga,\gp,0,0\) = 1,
\end{equation} 
yielding
\beqn
I_3^{D}\(\nu_1,\nu_2,\nu_3;Q_1^2,0,0,0,0,0\)\!\!
&=& \!\!(-1)^{\halfD} \(Q_1^2\)^{\halfD-\sigma} \nonumber\\
&\times & \!\!
\frac{\G{\halfD-\nu_1-\nu_2}\G{\halfD-\nu_1-\nu_3}\G{\sigma-\halfD}}
{\G{\nu_2}\G{\nu_3}\G{D-\sigma}},\phantom{aaaa}
\label{eq:tri1result}
\eeqn
which again agrees with the known result straightforwardly 
obtained using Feynman parameters.
Alternatively, we could have returned to the general
system~(\ref{eq:systrim1m2m3}),  
where, with only the $Q_1^2$ scale ($q_2=q_3=m_1=m_2=m_3=0$),  we 
would have had $(n+q+m)=4$ summation variables and $(n+1)=4$ constraints. 

\end{itemize}

\subsection{Two massive propagators and one off-shell leg}
\label{subsec:2mass}

We now turn to triangle integrals with two internal mass scales and one
external scale.  These have $(n+q+m)= 6$ summation variables and $(n+1)=4$
constraints and are therefore described by double sums.

\subsubsection[$M_1=0,~ Q_2^2=Q_3^2=0$]
{$\boldsymbol{M_1=0,~ Q_2^2=Q_3^2=0}$}
In this case, the system of constraints is obtained by setting
$m_1=q_2=q_3=0$ in Eq.~(\ref{eq:systrim1m2m3}).  The first constraint is
simply $p_1 = -\nu_1$ and there are only 8 solutions for the system.
As usual, each
solution is inserted into Eq.~(\ref{eq:trigen}) and treated accordingly to
the procedure of Sec~\ref{sec:general}.\\
The solutions can be grouped either by studying the physical thresholds of
the integral (or the convergence properties of the series) or by considering
the system, as in Sec.~\ref{sec:sys_constr}.\\
The threshold for the production of two massive propagators on-shell,
$\sqrt{Q_1^2} = M_2+M_3$, becomes evident when we inspect the hypergeometric
functions: four solutions are convergent above threshold,
$\sqrt{Q_1^2} > \sqrt{M_2^2}+\sqrt{M_3^2}$, while the other solutions equally
populate the regions $\sqrt{M_3^2} > \sqrt{M_2^2}+\sqrt{Q_1^2}$ and
$\sqrt{M_2^2} > \sqrt{M_3^2}+\sqrt{Q_1^2}$.

Consideration of the system reveals that the first group of solutions are
pairs from the set $\{m_2,m_3,p_2,p_3\}$ and the other groups formed are from
the sets $\{q_1,p_3,m_2\}$ and $\{q_1,p_2,m_3\}$ respectively.
The apparent overlap between the groups, solutions formed from the pairs
$\{m_2,p_3\}$ and $\{m_3,p_2\}$ are excluded by the system.  In each case, we
identify Appell's $F_4$ function, whose convergence properties match onto the
anticipated regions.

We find:
\begin{eqnarray}
\label{eq:sol_tri_q1m2m3}
\lefteqn{{\rm If\ } \sqrt{Q_1^2} > \sqrt{M_2^2}+\sqrt{M_3^2} \hspace{1.5cm}} \nonumber\\
\lefteqn{I_3^D(\nu_1,\nu_2,\nu_3;Q_1^2,0,0,0,M_2^2,M_3^2)=
I_3^{\{ m_2,m_3\}}+I_3^{\{ m_2,p_2\}} 
+I_3^{\{ m_3,p_3\}} +I_3^{\{ p_2,p_3\}} 
}\nonumber \\
&=& (-1)^{\halfD}   
\left(Q_1^2\right)^{\halfD-\nu_1-\nu_2-\nu_3}
\frac{\G{\nu_1+\nu_2+\nu_3-\halfD}\G{\halfD-\nu_1-\nu_2}\G{\halfD-\nu_1-\nu_3}}
{\G{\nu_2}\G{\nu_3}\G{D-\nu_1-\nu_2-\nu_3}} \nonumber \\
&& \times ~~~{F_4}\left(1+\sigma-D,\sigma-\halfD,
1+\nu_1+\nu_2-\halfD,1+\nu_1+\nu_3-\halfD,
\frac{M_2^2}{Q_1^2},\frac{M_3^2}{Q_1^2}\right) \nonumber \\
&+&  (-1)^{\halfD}\left(Q_1^2\right)^{-\nu_2}\left(-M_3^2\right)^{\halfD-\nu_1-\nu_3}
\frac{\G{\nu_1+\nu_3-\halfD}\G{\halfD-\nu_1-\nu_2}}
{\G{\nu_3}\G{\halfD-\nu_2}} \nonumber \\
&& \times ~~~{F_4}\left(1+\nu_2-\halfD,\nu_2,
1+\nu_1+\nu_2-\halfD,1+\halfD-\nu_1-\nu_3,
\frac{M_2^2}{Q_1^2},\frac{M_3^2}{Q_1^2}\right) \nonumber \\
&+&  (-1)^{\halfD}\left(Q_1^2\right)^{-\nu_3}\left(-M_2^2\right)^{\halfD-\nu_1-\nu_2}
\frac{\G{\nu_1+\nu_2-\halfD}\G{\halfD-\nu_1-\nu_3}}
{\G{\nu_2}\G{\halfD-\nu_3}} \nonumber \\
&& \times ~~~{F_4}\left(1+\nu_3-\halfD,\nu_3,
1+\halfD-\nu_1-\nu_2,1+\nu_1+\nu_3-\halfD,
\frac{M_2^2}{Q_1^2},\frac{M_3^2}{Q_1^2}\right) \nonumber \\
&+&  (-1)^{\halfD}\left(Q_1^2\right)^{\nu_1-\halfD}\left(-M_2^2\right)^{\halfD-\nu_1-\nu_2}\left(-M_3^2\right)^{\halfD-\nu_1-\nu_3}
\nonumber \\
&& \times
~~~\frac{\G{\nu_1+\nu_2-\halfD}\G{\nu_1+\nu_3-\halfD}\G{\halfD-\nu_1}}
{\G{\nu_1}\G{\nu_2}\G{\nu_3}} \nonumber \\
&& \times
~~~{F_4}\left(1-\nu_1,\halfD-\nu_1,1+\halfD-\nu_1-\nu_2,1+\halfD-\nu_1-\nu_3,
\frac{M_2^2}{Q_1^2},\frac{M_3^2}{Q_1^2}\right), \\ \nonumber
%%%%%%%%%%%%%%%%%%%%%%%%%%%%%%%%%%%%
&&\\
\label{eq:sol_tri_m3q1m2}
\lefteqn{{\rm if\ } \sqrt{M_3^2} > \sqrt{Q_1^2}+\sqrt{M_2^2} \hspace{1.5cm}} \nonumber\\
\lefteqn{I_3^D(\nu_1,\nu_2,\nu_3;Q_1^2,0,0,0,M_2^2,M_3^2)=
I_3^{\{ m_2,q_1\}}+I_3^{\{ p_3,q_1\}}
}\nonumber \\
&=& (-1)^{\halfD}  
\left(-M_3^2\right)^{\halfD-\nu_1-\nu_2-\nu_3}
\frac{\G{\nu_1+\nu_2+\nu_3-\halfD}\G{\halfD-\nu_1-\nu_2}}
{\G{\nu_3}\G{\halfD}} \nonumber \\
&& \times ~~~{F_4}\left(\nu_1+\nu_2+\nu_3-\halfD,\nu_2,
1+\nu_1+\nu_2-\halfD,\halfD,\frac{M_2^2}{M_3^2},
\frac{Q_1^2}{M_3^2}\right) \nonumber \\
&+& (-1)^{\halfD} \left(-M_2^2\right)^{\halfD-\nu_1-\nu_2}\left(-M_3^2\right)^{-\nu_3}
\frac{\G{\nu_1+\nu_2-\halfD}\G{\halfD-\nu_1}}
{\G{\nu_2}\G{\halfD}} \nonumber \\
&& \times
~~~{F_4}\left(\halfD-\nu_1,\nu_3,1+\halfD-\nu_1-\nu_2,\halfD,
\frac{M_2^2}{M_3^2},\frac{Q_1^2}{M_3^2}\right),
\end{eqnarray}
while the result for $\sqrt{M_2^2} > \sqrt{M_3^2}+\sqrt{Q_1^2}$ is obtained
by the exchanges $M_2 \leftrightarrow M_3$, $\nu_2 \leftrightarrow \nu_3$ in
Eq.~(\ref{eq:sol_tri_m3q1m2}).

We can check that these expressions are valid in certain limits.  
\paragraph{Checks}

\begin{itemize}
\item[-] {\bf The $\boldsymbol{\nu_1 \,\to\, 0}$ limit:
$\boldsymbol{I_3^D\(0,\nu_2,\nu_3;Q_1^2,0,0,0,M_2^2,M_3^2\)}$ }\\ Pinching
out the first propagator, the first three terms in (\ref{eq:sol_tri_q1m2m3})
and both terms in (\ref{eq:sol_tri_m3q1m2}) survive, yielding the general
bubble integral of Eq.~(\ref{eq:i2d_groups}), in the respective kinematic
regions,
\begin{equation}
I_3^D\(0,\nu_2,\nu_3;Q_1^2,0,0,0,M_2^2,M_3^2\)
=I_2^D\(\nu_2,\nu_3;Q_1^2,M_2^2,M_3^2\).
\end{equation}

\item[-] {\bf The $\boldsymbol{\nu_2 \,\to\, 0}$ limit:
$\boldsymbol{I_3^D\(\nu_1,0,\nu_3;Q_1^2,0,0,0,M_2^2,M_3^2\)}$ }\\ 
This limit should produce a self-energy integral with no external momentum
and a single internal mass $M_3$.  This is indeed the case: only the second
term in (\ref{eq:sol_tri_q1m2m3}) and the first term of
(\ref{eq:sol_tri_m3q1m2}) survive, each yielding the same result of
Eq.~(\ref{eq:sol_bub_q1m2to0}).

\item[-] {\bf The $\boldsymbol{M_2 \,\to\, 0}$ limit:
$\boldsymbol{I_3^D\(\nu_1,\nu_2,\nu_3;Q_1^2,0,0,0,0,M_3^2\)}$ }\\ 
Here the real production threshold occurs at $Q_1^2 = M_3^2$ and  
in this limit, Eq.~(\ref{eq:sol_tri_q1m2m3}) provides the $Q_1^2 > M_3^2$
result, while Eq.~(\ref{eq:sol_tri_m3q1m2}) gives the expression for
$M_3^2 > Q_1^2$.    
The Appell functions again collapse to form Gaussian hypergeometric
functions (see Eq.~(\ref{eq:F4_xtozero})), and we find:
\begin{eqnarray}
\label{eq:sol_tri_q1m3}
\lefteqn{{\rm If\ } \sqrt{Q_1^2} > \sqrt{M_3^2} \hspace{1.5cm}} \nonumber\\
\lefteqn{I_3^D(\nu_1,\nu_2,\nu_3;Q_1^2,0,0,0,0,M_3^2)=
I_3^{\{m_3 \}}+I_3^{\{p_3 \}}
}\nonumber \\
&=& (-1)^{\halfD}  
\left(Q_1^2\right)^{\halfD-\nu_1-\nu_2-\nu_3}
\frac{\G{\sigma-\halfD}\G{\halfD-\nu_1-\nu_2}\G{\halfD-\nu_1-\nu_3}}
{\G{\nu_2}\G{\nu_3}\G{D-\sigma}} \nonumber \\
&& \times ~~~{_2F_1}\left(1+\sigma-D,\sigma-\halfD,
1+\nu_1+\nu_3-\halfD,
\frac{M_3^2}{Q_1^2}\right) \nonumber \\
&+& (-1)^{\halfD} \left(Q_1^2\right)^{-\nu_2}\left(-M_3^2\right)^{\halfD-\nu_1-\nu_3}
\frac{\G{\nu_1+\nu_3-\halfD}\G{\halfD-\nu_1-\nu_2}}
{\G{\nu_3}\G{\halfD-\nu_2}} \nonumber \\
&& \times ~~~{_2F_1}\left(\nu_2,1+\nu_2-\halfD,
1+\halfD-\nu_1-\nu_3,
\frac{M_3^2}{Q_1^2}\right), \\ \nonumber
%%%%%%%%%%%%%%%%%%%%%%%%%%%%%%%%%%%
&&\\
\label{eq:sol_tri_m3q1}
\lefteqn{{\rm if\ } \sqrt{M_3^2} > \sqrt{Q_1^2} \hspace{1.5cm}} \nonumber\\
\lefteqn{I_3^D(\nu_1,\nu_2,\nu_3;Q_1^2,0,0,0,0,M_3^2)=
I_3^{\{q_1 \}}}\nonumber \\
&=& (-1)^{\halfD}
\left(-M_3^2\right)^{\halfD-\nu_1-\nu_2-\nu_3}
\frac{\G{\sigma-\halfD}\G{\halfD-\nu_1-\nu_2}}
{\G{\nu_3}\G{\halfD}} \,\f21\left(\nu_2,\sigma-\halfD,
\halfD,
\frac{Q_1^2}{M_3^2}\right). \nonumber \\
\end{eqnarray}
This latter result agrees with that obtained by taking the limit $Q_2^2 \,\to\, 0$,
$Q_3^2 \,\to\, 0$ in the general result given by Boos and Davydychev \cite{BD}
for a triangle loop integral with a single massive propagator.

\end{itemize}

\subsubsection[$M_3=0,\  Q_2^2=Q_3^2=0$]
{$\boldsymbol{M_3=0,\  Q_2^2=Q_3^2=0}$}
\label{sec:m3=0}
We now consider the triangle graph where $M_3=0$ and $Q_2^2=Q_3^2=0$.
Although this graph is not usually present in Standard Model processes, the
analysis of this graph turns out to be rather more subtle than the preceding
triangle integrals and we will therefore describe it in more detail.
Inspection of the singularities present in the loop integral via the Landau
equations reveals that threshold singularities occur at $M_2^2 = Q_1^2 +
M_1^2$.  We expect that this equality will provide the appropriate boundaries
of regions of convergence when considering the convergence properties of the
generalised hypergeometric functions. Furthermore, since the convergence
properties of these functions only depend on the absolute value of ratios of
scales, we expect that the reflections, $M_1^2 + M_2^2 = Q_1^2$ and $M_1^2 =
M_2^2 + Q_1^2$, will also form boundaries in the large $Q_1^2$ and $M_1^2$
regions, respectively. We also expect that, in certain limits, the solutions
match onto the kinematic regions relevant for simpler integrals.  For
example, in the limit $M_1 \,\to\, 0$, the discussion of the previous section
informs us that the solutions divide according to whether or not $Q_1^2 >
M_2^2$.  Similarly as $Q_1^2 \,\to\, 0$, there should be a threshold at $M_1 =
M_2$.

\begin{figure}
\begin{center}
~\epsfig{file=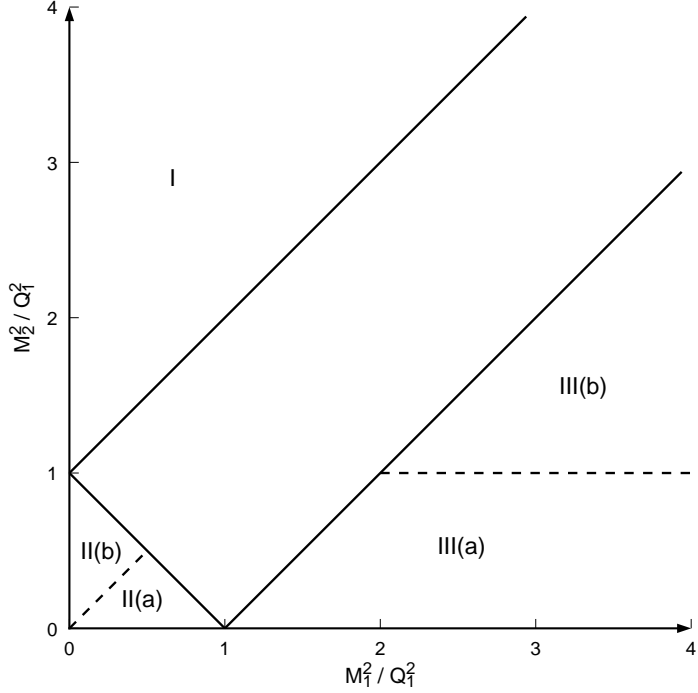,height=10cm}
\end{center}
\caption[]{The kinematic regions for the one-loop triangle with 
$Q_2^2 = Q_3^2 = M_3^2=0$.  The solid line shows the threshold in the Landau
surface at $M_2^2 = Q_1^2 + M_1^2$, together with the reflections $M_1^2 +
M_2^2 = Q_1^2$ and $M_1^2 = M_2^2 + Q_1^2$.  The reflections are relevant for
the convergence properties of the hypergeometric functions which only involve
the absolute values of ratios of the scales.  The dashed lines show the
boundaries $M_1^2 = M_2^2$ and $M_2^2 = Q_1^2$.}
\label{fig:region}
\end{figure}
In anticipation, we therefore divide the kinematic regions up as follows:
\begin{equation}
\label{eq:regions}
\begin{array}{ll}
{\rm region\ I:} & \displaystyle{M_2^2 > Q_1^2+M_1^2, }
\nonumber\vspace{1mm} \\  
{\rm region\ II(a):} &  \displaystyle{Q_1^2 > M_1^2+M_2^2 {\rm\ and\ } M_1^2
> M_2^2,} 
\nonumber\vspace{1mm}  \\  
{\rm region\ II(b):} &  \displaystyle{Q_1^2 > M_1^2+M_2^2 {\rm\ and\ } M_2^2
> M_1^2,}\vspace{1mm}  \\
{\rm region\ III(a):} & \displaystyle{ M_1^2 > Q_1^2+M_2^2 {\rm\ and\ }
Q_1^2 > M_2^2,} 
\nonumber\vspace{1mm}  \\  
{\rm region\ III(b):} & \displaystyle{ M_1^2 > Q_1^2+M_2^2 {\rm\ and\ } M_2^2
> Q_1^2,} 
\nonumber 
\end{array}
\end{equation}
as shown in Fig.~\ref{fig:region}.

With this set of scales, the system is given by Eq.~(\ref{eq:systrim1m2m3})
with $m_3 = q_2=q_3=0$.  As usual, we construct the solutions by solving the
system, inserting the solutions into Eq.~(\ref{eq:trigen}) and following the
procedure outlined in Sec~\ref{sec:general}.  Labelling each solution by the
summation variables and using the definitions of the hypergeometric functions
of Sec.~\ref{subsec:series}, we find:
\begin{eqnarray}
\label{eq:triangle_sols}
%%%% 3
I_3^{\{m_1,q_1\}} &=& (-1)^\halfD \left(-M_2^2\right)^{\halfD-\nu_1-\nu_2-\nu_3}
\frac{\G{\nu_1+\nu_2+\nu_3-\halfD}\G{\halfD-\nu_1-\nu_3}}
{\G{\nu_2}\G{\halfD}} \nonumber \\
&& \times
~~{F_2}\left(\nu_1+\nu_2+\nu_3-\halfD,\nu_1,\nu_3,1+\nu_1+\nu_3-\halfD,\halfD,
\frac{M_1^2}{M_2^2},
\frac{Q_1^2}{M_2^2}\right), \nonumber \\
%%%% 11
I_3^{\{p_2,q_1\}}&=& (-1)^\halfD \left(-M_1^2\right)^{\halfD-\nu_1-\nu_3}\left(-M_2^2\right)^{-\nu_2}
\frac{\G{\nu_1+\nu_3-\halfD}\G{\halfD-\nu_3}}
{\G{\nu_1}\G{\halfD}} \nonumber \\
&& \times
~~{F_2}\left(\nu_2,\halfD-\nu_3,\nu_3,1+\halfD-\nu_1-\nu_3,\halfD,
\frac{M_1^2}{M_2^2},\frac{Q_1^2}{M_2^2}\right), \nonumber \\
%%%% 1
I_3^{\{m_1,m_2\}} &=& (-1)^\halfD \left(Q_1^2\right)^{\halfD-\nu_1-\nu_2-\nu_3}
\frac{\G{\nu_1+\nu_2+\nu_3-\halfD}\G{\halfD-\nu_1-\nu_2}\G{\halfD-\nu_1-\nu_3}}
{\G{\nu_2}\G{\nu_3}\G{D-\nu_1-\nu_2-\nu_3}} \nonumber \\
&& \times ~~{S_1}\left(1+\sigma-D,\sigma-\halfD,
\nu_1,1+\nu_1+\nu_2-\halfD,1+\nu_1+\nu_3-\halfD,-\frac{M_1^2}{Q_1^2},
\frac{M_2^2}{Q_1^2}\right), \nonumber \\
%%%% 5
I_3^{\{m_2,p_2\}} &=& (-1)^\halfD \left(Q_1^2\right)^{-\nu_2}\left(-M_1^2\right)^{\halfD-\nu_1-\nu_3}
\frac{\G{\nu_1+\nu_3-\halfD}\G{\nu_3-\nu_2}\G{\halfD-\nu_3}}
{\G{\nu_1}\G{\nu_3}\G{\halfD-\nu_2}} \nonumber \\
&& \times
~~{S_1}\left(1+\nu_2-\halfD,\nu_2,\halfD-\nu_3,1+\nu_2-\nu_3,1+\halfD-\nu_1-\nu_3,
-\frac{M_1^2}{Q_1^2},\frac{M_2^2}{Q_1^2}\right), \nonumber \\
%%%% 6
I_3^{\{m_2,p_3\}} &=& (-1)^\halfD \left(Q_1^2\right)^{-\nu_3}\left(-M_1^2\right)^{\halfD-\nu_1-\nu_2}
\frac{\G{\nu_1+\nu_2-\halfD}\G{\nu_2-\nu_3}\G{\halfD-\nu_2}}
{\G{\nu_1}\G{\nu_2}\G{\halfD-\nu_3}} \nonumber \\
&& \times
~~{S_2}\left(\nu_2-\nu_3,\nu_1+\nu_2-\halfD,1+\nu_3-\halfD,\nu_3,
1+\nu_2-\halfD,
\frac{M_2^2}{M_1^2},\frac{M_1^2}{Q_1^2}\right), \nonumber \\
%%%% 8
I_3^{\{p_1,p_3\}} &=& (-1)^\halfD \left(Q_1^2\right)^{-\nu_3}\left(-M_1^2\right)^{-\nu_1}\left(-M_2^2\right)^{\halfD-\nu_2}
\frac{\G{\nu_2-\halfD}}
{\G{\nu_2}} \nonumber \\
&& \times
~~{F_3}\left(\nu_1,1+\nu_3-\halfD,\halfD-\nu_3,\nu_3,1+\halfD-\nu_2,
\frac{M_2^2}{M_1^2},\frac{M_2^2}{Q_1^2}\right), \nonumber \\
%%%% 2
I_3^{\{m_1,p_3\}} &=& (-1)^\halfD \left(Q_1^2\right)^{-\nu_3}\left(-M_2^2\right)^{\halfD-\nu_1-\nu_2}
\frac{\G{\nu_1+\nu_2-\halfD}\G{\halfD-\nu_1-\nu_3}}
{\G{\nu_2}\G{\halfD-\nu_3}} \nonumber \\
&& \times
~~{H_2}\left(\nu_1+\nu_2-\halfD,\nu_1,1+\nu_3-\halfD,\nu_3,1+\nu_1+\nu_3-\halfD,
\frac{M_1^2}{M_2^2},
-\frac{M_2^2}{Q_1^2}\right), \nonumber \\
%%%% 10
I_3^{\{p_2,p_3\}}&=& (-1)^\halfD \left(Q_1^2\right)^{-\nu_3}\left(-M_1^2\right)^{\halfD-\nu_1-\nu_3}\left(-M_2^2\right)^{\nu_3-\nu_2}
\frac{\G{\nu_1+\nu_3-\halfD}\G{\nu_2-\nu_3}}
{\G{\nu_1}\G{\nu_2}} \nonumber \\
&& \times
~~{H_2}\left(\nu_2-\nu_3,\halfD-\nu_3,1+\nu_3-\halfD,\nu_3,1+\halfD-\nu_1-\nu_3,
\frac{M_1^2}{M_2^2},-\frac{M_2^2}{Q_1^2}\right), \nonumber \\
%%%% 4
I_3^{\{m_2,p_1\}} &=& (-1)^\halfD \left(Q_1^2\right)^{\halfD-\nu_2-\nu_3}\left(-M_1^2\right)^{-\nu_1}
\frac{\G{\nu_2+\nu_3-\halfD}\G{\halfD-\nu_2}\G{\halfD-\nu_3}}
{\G{\nu_2}\G{\nu_3}\G{D-\nu_2-\nu_3}} \nonumber \\
&& \times ~~{S_2}\left(\nu_2+\nu_3-\halfD,1+\nu_2+\nu_3-D,
\nu_1,\halfD-\nu_3,1+\nu_2-\halfD,
\frac{M_2^2}{Q_1^2},\frac{Q_1^2}{M_1^2}\right), \nonumber \\
%%%% 7
I_3^{\{m_2,q_1\}} &=& (-1)^\halfD \left(-M_1^2\right)^{\halfD-\nu_1-\nu_2-\nu_3}
\frac{\G{\nu_1+\nu_2+\nu_3-\halfD}\G{\halfD-\nu_2-\nu_3}}
{\G{\nu_1}\G{\halfD}} \nonumber \\
&& \times
~~{S_1}\left(\nu_2,\nu_1+\nu_2+\nu_3-\halfD,\nu_3,1+\nu_2+\nu_3-\halfD,\halfD,
-\frac{Q_1^2}{M_1^2},\frac{M_2^2}{M_1^2}\right), \nonumber \\
%%%% 9
I_3^{\{p_1,q_1\}}&=& (-1)^\halfD \left(-M_1^2\right)^{-\nu_1}\left(-M_2^2\right)^{\halfD-\nu_2-\nu_3}
\frac{\G{\nu_2+\nu_3-\halfD}\G{\halfD-\nu_3}}
{\G{\nu_2}\G{\halfD}} \nonumber \\
&& \times
~~{H_2}\left(\nu_2+\nu_3-\halfD,\nu_3,\nu_1,\halfD-\nu_3,\halfD,
\frac{Q_1^2}{M_2^2},-\frac{M_2^2}{M_1^2}\right). 
\end{eqnarray}
We now need to study the convergence properties of these solutions.   
For example, by inspecting Table~\ref{tab:convergence} in
Appendix~\ref{subsec:series},  we see that 
the function $S_1(\ldots, x,y)$ is convergent when $|x| + |y| < 1$.
This implies that solution $I_3^{\{m_1,m_2\}}$ is 
convergent when
\begin{equation}
\left | \frac{-M_1^2}{Q_1^2} \right | +\left | \frac{M_2^2}{Q_1^2} \right | <
1, 
\end{equation} 
or, in other words, 
\begin{equation} 
M_1^2 + M_2^2 < Q_1^2,
\end{equation}
independently of whether $M_1$ is larger than $M_2$ or not.
This series therefore converges  in both regions II(a) and II(b).\\
On the other hand, $I_3^{\{m_1,p_3\}}$ converges when
\begin{equation} 
-\left | \frac{M_1^2}{M_2^2} \right | +\left | \frac{Q_1^2}{M_2^2} \right | > 1
\qquad {\rm and} \qquad \left | \frac{M_1^2}{M_2^2} \right |< 1
\qquad {\rm and} \qquad \left | \frac{M_2^2}{Q_1^2} \right |< 1,
\end{equation}
or, alternatively,
\begin{equation} 
M_1^2 + M_2^2 < Q_1^2 \qquad {\rm and} \qquad M_2^2 > M_1^2,
\end{equation}
which corresponds to region II(b) only.

Applying the convergence criteria to each of the eleven solutions, we find
that they are distributed as follows:
\begin{eqnarray}
{\lefteqn{\rm in~region~I}\hspace{6cm}} 
\nonumber \\
\label{eq:sol_tri_I}
I_3^D(\nu_1,\nu_2,\nu_3;Q_1^2,0,0,M_1^2,M_2^2,0)&=&
I_3^{\{ m_1,q_1 \}}+I_3^{\{ p_2,q_1 \}},
\\
{\lefteqn{\rm in~region~II(a)}\hspace{6cm}} 
\nonumber \\
\label{eq:sol_tri_IIa}
I_3^D(\nu_1,\nu_2,\nu_3;Q_1^2,0,0,M_1^2,M_2^2,0)&=&
I_3^{\{ m_1,m_2 \}}+I_3^{\{ m_2,p_2 \}}
+I_3^{\{ m_2,p_3 \}}+I_3^{\{ p_1,p_3 \}},
\\
{\lefteqn{\rm in~region~II(b)}\hspace{6cm}} 
\nonumber \\
\label{eq:sol_tri_IIb}
I_3^D(\nu_1,\nu_2,\nu_3;Q_1^2,0,0,M_1^2,M_2^2,0)&=&
I_3^{\{ m_1,m_2 \}}+I_3^{\{ m_2,p_2 \}}
+I_3^{\{ m_1,p_3 \}}+I_3^{\{ p_2,p_3 \}},
\\
{\lefteqn{\rm in~region~III(a)}\hspace{6cm}} 
\nonumber \\
\label{eq:sol_tri_IIIa}
I_3^D(\nu_1,\nu_2,\nu_3;Q_1^2,0,0,M_1^2,M_2^2,0)&=&
I_3^{\{ m_2,p_1 \}}+I_3^{\{ m_2,q_1 \}}
+I_3^{\{ p_1,p_3 \}},
\\
{\lefteqn{\rm in~region~III(b)}\hspace{6cm}} 
\nonumber \\
\label{eq:sol_tri_IIIb}
I_3^D(\nu_1,\nu_2,\nu_3;Q_1^2,0,0,M_1^2,M_2^2,0)&=&
I_3^{\{ m_2,q_1 \}}+I_3^{\{ p_1,q_1 \}}.\hspace{6cm}
\end{eqnarray}
We see that in region II(a), two of the solutions
$\(I_3^{\{m_2,p_2\}}\right.$ and 
$\left. I_3^{\{m_2,p_3\}}\)$ contain dangerous $\Gamma$ functions when
$\nu_2 = 
\nu_3$.  These divergences usually indicate the region of a logarithmic
analytic continuation and can be regulated by letting $\nu_2 = \nu_3 +
\delta$, canceling the divergence, and then setting $\delta \,\to\, 0$.
Similarly, the two divergent contributions in region II(b)
$\(I_3^{\{m_2,p_2\}}\right.$ and 
$\left. I_3^{\{p_2,p_3\}}\)$ also cancel in
this limit.

We have performed several checks of the correctness of this assignment into
groups.
\paragraph{Checks}
\begin{itemize}

\item[-]  {\bf Analytic continuation}\\
Applying the analytic continuation formulae given in Appendix~\ref{sec:app},
we can see that the solutions are connected to each other.  For example,
applying Eq.~(\ref{eq:f2_anal_cont}) to the $F_2$ functions in region I
produces the $H_2$ and $S_1$ functions of region II(b).  Similarly,
Eqs.~(\ref{eq:s1_anal_cont}) and~(\ref{eq:h2_anal_cont2}) transform the $S_1$
and $H_2$ solutions of region III(b) into the $F_2$ functions of region I.

\item[-] {\bf The $\boldsymbol{\nu_1=\nu_2=\nu_3=1}$ limit:
$\boldsymbol{I_3^D(1,1,1;Q_1^2,0,0,M_1^2,M_2^2,0)}$ }\\ 
All the groups give the correct answer when all the propagators are set equal
to one.

As an example, we consider region II(b), so that we can explicitly show the
cancellation of the $\de$ poles.  We fix $\nu_1=\nu_3=1$, $\nu_2=1+\delta$
and $D = 4-2\ep$.  For these choices of the parameters, the hypergeometric
functions simplify using the identities given in Sec.~\ref{app:reduction}, and
we find
\begin{eqnarray}
%%%% 1
I_3^{\{m_1,m_2\}} &=&
-\frac{N_\ep}{\ep^2} \frac{\G{1-\ep}^2}{\G{1-2\ep}}
\(\frac{Q_1^2}{Q_1^2-M_2^2}\)^{2\ep}  
\f21\(1,2\,\epsilon,1+\epsilon,\frac{M_1^2}{M_2^2-Q_1^2}\)
\label{eq:I3m1m2}\\
%%%%%%%%%%%%%%%%%%%%%%%
I_3^{\{m_2,p_2\}} &=&
\frac{N_\ep}{\ep} \frac{\G{1-\ep}}{\G{1-\ep-\de}} \frac{\G{1-\de}}{\de}
\frac{\(Q_1^2\)^{2\ep}}{\(-M_1^2\)^\ep}
\left(Q_1^2+M_1^2-M_2^2\right)^{-\epsilon-\de}
\label{eq:I3m2p2}\\
%%%%%%%%%%%%%%%%%%%%%%%%%%%%
I_3^{\{m_1,p_3\}} &=& 
\frac{N_\ep}{\ep^2} \frac{M_2^2}{M_2^2-M_1^2} \(\frac{Q_1^2}{-M_2^2}\)^\ep
F_2\(1,1,\ep,\ep+1,1-\ep,\frac{M_1^2}{M_1^2-M_2^2},\frac{M_2^2}{Q_1^2}\)
\phantom{aaaa}
\label{eq:I3m1p3} \\
%%%%%%%%%%%%%%%%%%%%%%%%%%%%
I_3^{\{p_2,p_3\}} &=& 
-\frac{N_\ep}{\ep} \frac{(-1)^{-\de}}{\de\, \( M_2^2-M_1^2\)^\de}
\(\frac{Q_1^2}{-M_1^2}\)^\ep
\f21\(1,\ep,1-\de,\frac{M_2^2-M_1^2}{Q_1^2}\),
\label{eq:I3p2p3}
\end{eqnarray}
where we have defined
\begin{equation}
  N_\ep = \G{1+\ep} \(-Q_1^2\)^{-1-\ep}.
\end{equation}
Using Eq.~(\ref{eq:f21_anal_1-1/z}), we can rewrite $I_3^{\{p_2,p_3\}}$ in the
following way
\beqn
I_3^{\{p_2,p_3\}}\!\!\!\! &=& \!\!
-\frac{N_\ep}{\ep^2} \(\frac{Q_1^2}{-M_1^2}\)^\ep \frac{Q_1^2}{M_2^2-M_1^2}
~\f21\(1,1,1+\ep,\frac{Q_1^2+M_1^2-M_2^2}{M_1^2-M_2^2}\) 
\nonumber\\
&-&\!\!\frac{N_\ep}{(\de+\ep)}  \frac{\G{1+\ep+\de}}{\G{1+\ep} }
 \frac{ (-1)^{-\de} \G{1-\de}}{\de} \frac{\(Q_1^2\)^{2\ep}}{\(-M_1^2\)^\ep}
\(Q_1^2+M_1^2-M_2^2\)^{-\ep-\de},\phantom{aaa}
\label{eq:I3p2p3_del}
\eeqn
where we have safely put $\de=0$ in the first line, since this is a finite
quantity in $\de$.

We see that the poles in $\de$ clearly cancel between $I_3^{\{m_2,p_2\}}$ and
$I_3^{\{p_2,p_3\}}$ (see Eqs.~(\ref{eq:I3m2p2}) and~(\ref{eq:I3p2p3_del})),
leaving a finite remainder that is straightforwardly obtained by Taylor
expansion about $\de = 0$.

Up to now, we have not required $\ep$ to be small and 
expressions~(\ref{eq:I3m1m2})--(\ref{eq:I3p2p3_del}) are valid in arbitrary
dimension $D$. 

If we make the usual expansion for $\ep\,\to\, 0$ we recover the result
\begin{equation}
\label{eq:I3_ep_expans}
I_3^D(1,1,1;Q_1^2,0,0,M_1^2,M_2^2,0) = \! N_\ep\! \left[
\li{ \frac{Q_1^2+M_1^2-M_2^2}{M_1^2}}- \li{1-\frac{M_2^2}{M_1^2}}\right]
+\ord{\ep}, 
\end{equation} 
where we have used a series expansion for the integral representation of the
functions $\f21$ and $F_2$, given in Eqs.~(\ref{eq:f21_integral})
and~(\ref{eq:f2_integral}), and where $Q_1^2 \,\to \, Q_1^2 +i0$, to recover
the correct prescription in the Feynman integrals.
We describe the details of the $\ep$ expansion in
Appendix~\ref{subsec:ep_expansion}. 
Expression~(\ref{eq:I3_ep_expans}) is finite in $\ep$, as it should be,
having no soft or collinear singularities, despite the fact
that the individual contributions contain poles in $\ep$.

\item[-] {\bf The $\boldsymbol{\nu_1 \, \to \, 0}$ limit:
$\boldsymbol{I_3^D\(0,\nu_2,\nu_3;Q_1^2,0,0,M_1^2,M_2^2,0\)}$ }\\ 
If we set $\nu_1 \,\to\, 0$, we produce a one-mass ($M_2$) bubble
integral with external scale $Q_1^2$ and internal propagators raised to the
powers $\nu_2$ and $\nu_3$. \\ 
In the different regions~(\ref{eq:regions}) and for the different groups of
Eqs.~(\ref{eq:sol_tri_I})--(\ref{eq:sol_tri_IIIb}), we have
\begin{equation}
\label{eq:tri_v1equal0}
\begin{array}{llll}
{\rm I:} & \displaystyle{M_2^2 > Q_1^2 {\rm\ and\ } M_2^2>M_1^2} &
\Longrightarrow &
\displaystyle{\left. I_3^D \right|_{\nu_1=0} =
\left. I_3^{\{ m_1,q_1 \}}\right|_{\nu_1=0}
\phantom{\Biggl]}
} 
\nonumber \\  
{\rm  II(a):} &  \displaystyle{Q_1^2 > M_1^2 > M_2^2} &
\Longrightarrow &
\displaystyle{\left.I_3^D \right|_{\nu_1=0} =
\left. I_3^{\{ m_1,m_2 \}}\right|_{\nu_1=0} + \left. I_3^{\{ p_1,p_3 \}}\right|_{\nu_1=0}
\phantom{\Biggl]}
}
\nonumber \\  
{\rm  II(b):} &  \displaystyle{Q_1^2 > M_2^2 > M_1^2} &
\Longrightarrow &
\displaystyle{\left.I_3^D \right|_{\nu_1=0}=
\left. I_3^{\{ m_1,m_2 \}}\right|_{\nu_1=0} + \left. I_3^{\{ m_1,p_3 \}}\right|_{\nu_1=0}
\phantom{\Biggl]}
}
\nonumber \\
{\rm  III(a):} & \displaystyle{ M_1^2 > Q_1^2 > M_2^2}  &
\Longrightarrow &
\displaystyle{\left.I_3^D \right|_{\nu_1=0}=
\left. I_3^{\{ m_2,p_1 \}}\right|_{\nu_1=0}+\left. I_3^{\{ p_1,p_3 \}}\right|_{\nu_1=0}
\phantom{\Biggl]}
}
\nonumber \\  
{\rm  III(b):} & \displaystyle{M_1^2 > M_2^2
> Q_1^2 } &
\Longrightarrow &
\displaystyle{\left.I_3^D \right|_{\nu_1=0}=
\left. I_3^{\{ p_1,q_1 \}}\right|_{\nu_1=0},
\phantom{\Biggl]}
}
\nonumber 
\end{array}
\end{equation}
where we have used the shorthand notation
\begin{equation}
\left. I_3^D \right|_{\nu_1=0} =
I_3^D\(0,\nu_2,\nu_3;Q_1^2,0,0,M_1^2,M_2^2,0\),
\end{equation}
and where the missing terms have been killed by the $\G{0}$ in the
denominator.

It is straightforward to evaluate the different solutions when $\nu_1=0$.
In fact, taking $I_3^{\{ m_1,q_1 \}}$ as example, we can use the reduction
formula 
\begin{equation}
F_2\(\al,0,\bp,\ga,\gp,x,y\) = \f21\(\al,\bp,\gp,y\),
\end{equation}
to recover
\begin{equation}
I_3^D(0,\nu_2,\nu_3;Q_1^2,0,0,M_1^2,M_2^2,0) = I_2^D(\nu_2,\nu_3;Q_1^2,M_2^2,0)
\end{equation}
in region I, that is Eq.~(\ref{eq:sol_bub_m1}). The same thing happens to the
solution in region III(b).

The other part of the bubble integral valid when $Q_1^2 >
M_2^2$ is produced by the other solutions in the
region II(a), II(b) and III(a), and agrees with Eq.~(\ref{eq:sol_bub_q1}).

\item[-] {\bf The $\boldsymbol{\nu_3 \, \to \, 0}$ limit:
$\boldsymbol{I_3^D\(\nu_1,\nu_2,0;Q_1^2,0,0,M_1^2,M_2^2,0\)}$ }\\ 
Likewise, we can set $\nu_3 \,\to \, 0$ producing a two-mass bubble ($M_1$
and $M_2$) with external scale $Q_1^2 = 0$, for which the result is given in
Eq.~(\ref{eq:sol_bub_q1to0}).  
We could repeat the reasoning made for the previous case, and build a table
of surviving solutions, analogous to~(\ref{eq:tri_v1equal0}).

For example, the two terms in Eq.~(\ref{eq:sol_tri_I}) collapse to form the
correct Gauss' hypergeometric functions when $M_2^2 > M_1^2$ and $M_2^2 >
Q_1^2$.  Similarly, the result when $Q_1^2 > M_2^2> M_1^2$ is produced by the
third and fourth term of Eq.~(\ref{eq:sol_tri_IIb}), for region II(b).

\item[-] {\bf The $\boldsymbol{M_1 \, \to \, 0}$ limit:
$\boldsymbol{I_3^D\(\nu_1,\nu_2,\nu_3;Q_1^2,0,0,0,M_2^2,0\)}$ }\\ 
Here we expect to reproduce the result for the triangle integral given in
Eqs.~(\ref{eq:sol_tri_q1m3}) and ~(\ref{eq:sol_tri_m3q1}), with the
exchanges $M_3 \leftrightarrow M_2$ and $\nu_3 \leftrightarrow \nu_2$.
Clearly in regions II(a), III(a) and III(b), it is inappropriate to take this
limit, since $M_1^2 > M_2^2$.  In fact, if we just go ahead and apply the
limit blindly to the solutions for regions III(a) and III(b), we just obtain
zero.  \\
On the other hand, in regions I and II(b) it does make sense to send
$M_1 \,\to\, 0$ since $M_1$ is allowed to be the smallest scale present:
\begin{equation}
\label{eq:tri_M1equal0}
\begin{array}{llll}
{\rm I:} & \displaystyle{M_2^2 > Q_1^2 {\rm\ and\ } M_2^2>M_1^2} &
\Longrightarrow &
\displaystyle{\left. I_3^D \right|_{M_1=0} =
\left. I_3^{\{ m_1,q_1 \}}\right|_{M_1=0}
\phantom{\Biggl]}
} 
\nonumber \\  
{\rm  II(b):} &  \displaystyle{Q_1^2 > M_2^2 > M_1^2} &
\Longrightarrow &
\displaystyle{\left.I_3^D \right|_{M_1=0}=
\left. I_3^{\{ m_1,m_2 \}}\right|_{M_1=0} + \left. I_3^{\{ m_1,p_3
\}}\right|_{M_1=0},
\phantom{\Biggl]}
}
\nonumber 
\end{array}
\end{equation}
with the shorthand notation
\begin{equation}
\left. I_3^D \right|_{M_1=0} =
I_3^D\(\nu_1,\nu_2,\nu_3;Q_1^2,0,0,0,M_2^2,0\).
\end{equation}
Again the hypergeometric functions collapse to  Gauss' $\f21$ functions (see
Eqs.~(\ref{eq:F2_xtozero}), (\ref{eq:S1_xtozero}) and~(\ref{eq:H2_xtozero}))
\beqn
F_2\(\al,\bt,\bp,\ga,\gp,0,y\)  &=&  \f21\(\al,\bp,\gp,y\), \nonumber\\
S_1\(\al,\ap,\bt,\ga,\de,0,y\)  &=& \f21\(\al,\ap,\ga,y\),  \nonumber \\
H_2\(\al,\bt,\ga,\de,\ep,0,y\)  &=& \f21\(\ga,\de,1-\al,-y\),   \nonumber
\eeqn
and we recover, in region I, the result of Eq.~(\ref{eq:sol_tri_m3q1}), and
in region II(b), the expected result~(\ref{eq:sol_tri_q1m3}) for $M_2^2 <
Q_1^2$.

\item[-] {\bf The $\boldsymbol{M_2 \, \to \, 0}$ limit:
$\boldsymbol{I_3^D\(\nu_1,\nu_2,\nu_3;Q_1^2,0,0,M_1^2,0,0\)}$ }\\ 
Taking the limit $M_2 \,\to\, 0$ provides us with the integral relevant for
the exchange of a heavy particle in the decay into two light particles.

As usual, we could merely return to the system and, by setting $m_2=0$, solve
it afresh: there are now $(m+q+n) = 5$ variables and still $(n+1)=4$
constraints leaving five single-sum solutions.
However, it is simpler to take the $M_2 \, \to \,0$ limit in the appropriate
regions: II(a) and III(a), as can be seen from Eq.~(\ref{eq:regions}).

We then obtain
\begin{equation}
\label{eq:tri_m2equal0}
\begin{array}{llll}
{\rm  II(a):} &  \displaystyle{Q_1^2 > M_1^2 > M_2^2} &
\Longrightarrow & 
\displaystyle{\left.I_3^D \right|_{M_2=0} =
\left| I_3^{\{ m_1,m_2 \}}+
 I_3^{\{ m_2,p_2 \}}+
 I_3^{\{ m_2,p_3 \}}\right|_{M_2=0}
\phantom{\Biggl]}
}
\nonumber \\
{\rm  III(a):} & \displaystyle{ M_1^2 > Q_1^2 > M_2^2}  &
\Longrightarrow &
\displaystyle{\left.I_3^D \right|_{M_2=0}=
\left| I_3^{\{ m_2,p_1 \}}+
I_3^{\{ m_2,q_1 \}}\right|_{M_2=0},
\phantom{\Biggl]}
} \nonumber
\end{array}
\end{equation}
where 
\begin{equation}
\left. I_3^D \right|_{M_2=0} =
I_3^D\(\nu_1,\nu_2,\nu_3;Q_1^2,0,0,M_1^2,0,0\).
\end{equation}
The hypergeometric functions collapse to $\ftt$ functions, according
to Eqs.~(\ref{eq:S1_ytozero}) and~(\ref{eq:S2_xtozero})
\beqn
S_1\( \al,\ap,\bt,\ga,\de,x,0\) 
&=& {_3F_2}\(\al,\ap,\bt,\ga,\de,x\), \nonumber \\ 
S_2\(\al,\ap,\bt,\bp,\ga,0,y\) 
&=& {_3F_2}\(1-\ga,\bt,\bp,1-\al,1-\ap,-y\), \nonumber 
\eeqn
and we obtain:
\begin{eqnarray}
\label{eq:sol_tri_q1m1}
\lefteqn{{\rm if\ } Q_1^2 > M_1^2, {\rm \  \ region\ II(a)} 
\hspace{1.5cm}} \nonumber\\
\lefteqn{I_3^D(\nu_1,\nu_2,\nu_3;Q_1^2,0,0,M_1^2,0,0)=
I_3^{\{ m_1 \}}+I_3^{\{ p_2 \}}+I_3^{\{ p_3 \}}
}\nonumber \\
&=& (-1)^\halfD \left(Q_1^2\right)^{\halfD-\nu_1-\nu_2-\nu_3}
\frac{\G{\nu_1+\nu_2+\nu_3-\halfD}\G{\halfD-\nu_1-\nu_2}\G{\halfD-\nu_1-\nu_3}}
{\G{\nu_2}\G{\nu_3}\G{D-\nu_1-\nu_2-\nu_3}} \nonumber \\
&& \times ~~~{_3F_2}\left(\nu_1,1+\sigma-D,
\sigma-\halfD,1+\nu_1+\nu_2-\halfD,1+\nu_1+\nu_3-\halfD,
-\frac{M_1^2}{Q_1^2}\right) \nonumber \\
&+& (-1)^\halfD \left(Q_1^2\right)^{-\nu_2}\left(-M_1^2\right)^{\halfD-\nu_1-\nu_3}
\frac{\G{\nu_3-\nu_2}\G{\halfD-\nu_3}\G{\nu_1+\nu_3-\halfD}}
{\G{\nu_1}\G{\nu_3}\G{\halfD-\nu_2}} \nonumber \\
&& \times ~~~{_3F_2}\left(\nu_2,\halfD-\nu_3,1+\nu_2-\halfD,
1+\nu_2-\nu_3,1+\halfD-\nu_1-\nu_3,-\frac{M_1^2}{Q_1^2}\right) \nonumber \\
&+&  (-1)^\halfD\left(Q_1^2\right)^{-\nu_3}\left(-M_1^2\right)^{\halfD-\nu_1-\nu_2}
\frac{\G{\nu_2-\nu_3}\G{\halfD-\nu_2}\G{\nu_1+\nu_2-\halfD}}
{\G{\nu_1}\G{\nu_2}\G{\halfD-\nu_3}} \nonumber \\
&& \times ~~~{_3F_2}\left(\nu_3,\halfD-\nu_2,1+\nu_3-\halfD,
1+\nu_3-\nu_2,1+\halfD-\nu_1-\nu_2,-\frac{M_1^2}{Q_1^2}\right),\phantom{aaaaa}
\\
\label{eq:sol_tri_m1q1}
\lefteqn{{\rm if\ } M_1^2 > Q_1^2, {\rm \  \ region\ III(a)} 
 \hspace{1.5cm}} \nonumber\\
\lefteqn{I_3^D(\nu_1,\nu_2,\nu_3;Q_1^2,0,0,M_1^2,0,0)=
I_3^{\{ q_1 \}}+I_3^{\{ p_1 \}}
}\nonumber \\
&=&  (-1)^\halfD\left(-M_1^2\right)^{\halfD-\nu_1-\nu_2-\nu_3}
\frac{\G{\nu_1+\nu_2+\nu_3-\halfD}\G{\halfD-\nu_2-\nu_3}}
{\G{\nu_1}\G{\halfD}} \nonumber \\
&& \times ~~~{_3F_2}\left(\nu_2,\nu_3,\nu_1+\nu_2+\nu_3-\halfD,
1+\nu_3+\nu_2-\halfD,\halfD,-\frac{Q_1^2}{M_1^2}\right) \nonumber \\
&+& (-1)^\halfD \left(Q_1^2\right)^{\halfD-\nu_2-\nu_3}\left(-M_1^2\right)^{-\nu_1}
\frac{\G{\nu_2+\nu_3-\halfD}\G{\halfD-\nu_2}\G{\halfD-\nu_3}}
{\G{\nu_2}\G{\nu_3}\G{D-\nu_2-\nu_3}} \nonumber \\
&& \times ~~~{_3F_2}\left(\nu_1,\halfD-\nu_2,\halfD-\nu_3,
D-\nu_2-\nu_3,1+\halfD-\nu_2-\nu_3,-\frac{Q_1^2}{M_1^2}\right).
\end{eqnarray}

\end{itemize}

We can further check these results by setting one of the $\nu_i \,\to\, 0$ to
form bubble integrals or by taking one of the limits $M_1 \,\to\, 0$ or
$Q_1^2\, \to\, 0$.  In each case, we recover the correct results presented in
the earlier sections

\paragraph{Discussion of system of constraints} \mbox{} \newline
As we have already anticipated in Sec.~\ref{sec:sys_constr}, the homogeneous
counterpart of the system of constraints~(\ref{eq:systrim1m2m3}), with
$q_2=q_3=m_3=0$, is not uniquely solvable.
We can  build the table of signs for $p_i$, $q_i$ and $m_i$:
\begin{center}
\leavevmode
%\footnotesize
\begin{tabular}{c| c | c}
 \hline
$>0$  &  $<0$ & uncertain\\
\hline
$p_2, \; q_1, \; m_1$ &  $p_1, \; p_3, \; m_2$ & \\
$p_2, \; p_3, \; m_1 $ & $p_1, \; q_1$ & $ m_2$\\
$p_1, \; q_1$ &  $p_2, \; p_3, \;m_1$ &  $m_2$\\
$p_1, \; p_3, \; m_2$ &  $p_2, \; q_1, \; m_1$ & \\
      \hline
\end{tabular}
%\caption{}
\end{center}
but this does not determine the complete list of groups, even if some of
them (first and second row, corresponding to region I and region II(b)) are
correctly predicted by the table (if we arbitrarily assume that $m_2$ in the
second row has a positive sign).

\subsection[Equal-mass propagators: $M_1=M_2 = M_3= M$]
{Equal-mass propagators: $\boldsymbol{M_1=M_2 = M_3= M}$}

In the special case of $M_1 = M_2 =M_3=M$ we use the modified form given in
Sec.~\ref{subsec:equal}.  For the most general case of unequal off-shell
external legs, there are $(n+q+1) = 7$ summation variables and $(n+1) = 4$
constraints.  The template solution is easily obtained from
Eq.~(\ref{eq:sumequal}) and the system of constraints is given by
\begin{eqnarray}
q_2+q_3+p_1 &=& -\nu_1,\nonumber \\
q_1+q_3+p_2 &=& -\nu_2,\nonumber \\
q_1+q_2+p_3 &=& -\nu_3,\nonumber \\
p_1+p_2+p_3 +q_1+q_2+q_3&=& -\halfD +m.
\label{eq:systriequal}
\end{eqnarray}

\subsubsection[$Q_2^2 = Q_3^2 = 0$]
{$\boldsymbol{Q_2^2 = Q_3^2 = 0}$}
In this case, there are five summation variables ($p_1$, $p_2$, $p_3$, $q_1$
and $m$), and four constraints, and the system admits four solutions out of
the possible five. 

Based on two-particle cuts of the diagram, we see that there is a threshold
at $Q_1^2 = 4M^2$ corresponding to producing propagators 2 and 3 on-shell.  Of
the four solutions, three converge when $Q_1^2 > 4 M^2$, while the remaining
solution converges when $Q_1^2 < 4 M^2$.  Recalling
$\sigma=\nu_1+\nu_2+\nu_3$ (see Eq.~(\ref{eq:sigma})), we have:
\begin{eqnarray}
\label{eq:sol_tri_eqm3}
\lefteqn{{\rm if\ } Q_1^2 > 4 M^2 \hspace{1.5cm}} \nonumber\\
\lefteqn{I_3^D(\nu_1,\nu_2,\nu_3;Q_1^2,0,0,M^2,M^2,M^2)=
I_3^{\{ m \}}+I_3^{\{ p_2 \}}+I_3^{\{p_3 \}}}\nonumber \\
&=& (-1)^\halfD \left(Q_1^2\right)^{\halfD-\sigma}
\frac{\G{\sigma-\halfD}\G{\halfD-\nu_1-\nu_2}\G{\halfD-\nu_1-\nu_3}}
{\G{\nu_2}\G{\nu_3}\G{D-\sigma}} \nonumber \\
&\times & \! {_3F_2}\left(\sigma-\halfD,
1+\frac{1}{2}(\sigma-D),
\frac{1}{2}(1+\sigma-D),
1+\nu_1+\nu_2-\halfD,1+\nu_1+\nu_3-\halfD,
\frac{4M^2}{Q_1^2}\right) \nonumber \\
&+& (-1)^\halfD\left(Q_1^2\right)^{-\nu_2}(-M^2)^{\halfD-\nu_1-\nu_3}
\frac{\G{\nu_3-\nu_2}\G{\nu_1+\nu_3-\halfD}}
{\G{\nu_3}\G{\nu_1-\nu_2+\nu_3}} \nonumber \\
& \times & \! {_3F_2}\left(\nu_2,1+\frac{1}{2}(\nu_2-\nu_1-\nu_3),
\frac{1}{2}(1-\nu_1+\nu_2-\nu_3),
1+\nu_2-\nu_3,1+\halfD-\nu_1-\nu_3,
\frac{4M^2}{Q_1^2}\right) \nonumber \\
&+& (-1)^\halfD\left(Q_1^2\right)^{-\nu_3}(-M^2)^{\halfD-\nu_1-\nu_2}
\frac{\G{\nu_2-\nu_3}\G{\nu_1+\nu_2-\halfD}}
{\G{\nu_2}\G{\nu_1+\nu_2-\nu_3}} \nonumber \\
& \times & \! {_3F_2}\left(\nu_3,1+\frac{1}{2}(\nu_3-\nu_1-\nu_2),
\frac{1}{2}(1-\nu_1+\nu_3-\nu_2),
1+\nu_3-\nu_2,1+\halfD-\nu_1-\nu_2,
\frac{4M^2}{Q_1^2}\right),\nonumber \\
&&\\
\label{eq:sol_tri_eqm1}
\lefteqn{{\rm if\ } Q_1^2 < 4 M^2 \hspace{1.5cm}} \nonumber\\
\lefteqn{I_3^D(\nu_1,\nu_2,\nu_3;Q_1^2,0,0,M^2,M^2,M^2)=I_3^{\{ q_1 \}}}\nonumber \\
&=&(-1)^\halfD (-M^2)^{\halfD-\sigma}
\frac{\G{\sigma-\halfD}}
{\G{\sigma}} \  {_3F_2}\left(\nu_2,\nu_3,\sigma-\halfD,
\frac{\sigma}{2},\frac{1+\sigma}{2},
\frac{Q_1^2}{4M^2}\right),
\end{eqnarray}
where we have made use of formula~(\ref{eq:poch2n}).
This latter result agrees with that obtained by taking the limit $Q_2^2 \,
\to \, 0$, $Q_3^2 \,\to \,0$ in the general result given by Boos and Davydychev
\cite{BD} for a triangle loop integral with three off-shell legs and a single
mass $M$ running round the loop.  Equation~(\ref{eq:sol_tri_eqm3}) appears to
be a new result.  

We note that, taking the limit $\nu_1 \,\to \,0$ in
Eqs~(\ref{eq:sol_tri_eqm3}) and~(\ref{eq:sol_tri_eqm1}), we reproduce the
expected equal-mass bubble integral of Eqs.~(\ref{eq:sol_bub_eqm3})
and~(\ref{eq:sol_bub_eqm1})
\begin{equation}
I_3^D(0,\nu_2,\nu_3;Q_1^2,0,0,M^2,M^2,M^2)
=I_2^D(\nu_2,\nu_3;Q_1^2,M^2,M^2),
\end{equation}
while taking $M \,\to \, 0$, only the first term in
Eq.~(\ref{eq:sol_tri_eqm3}) survives, yielding Eq.~(\ref{eq:tri1result}).

However we observe that there are dangerous $\Gamma$ functions in
the second and third lines when $\nu_2 = \nu_3$. Therefore to
evaluate the integral when $\nu_2 = \nu_3 = \nu$, we introduce an
additional regulator $\delta$ such that $\nu_2=\nu+\delta$ and
$\nu_3 = \nu$.  As in the previous section, because the result does
not depend on $\delta$, the limit $\delta \, \to \, 0$ can be safely
taken after the singularities have been canceled.

%%%%%%%%%%%%%%%%%%%%%%%%%%%%%%%%%%%%%%%%%%%%%%%%% 
\section{Conclusions} 
\setcounter{equation}{0}
\label{sec:conc}
%%%%%%%%%%%%%%%%%%%%%%%%%%%%%%%%%%%%%%%%%%%%%%%%% 

Finally let us summarize what we have accomplished in this paper. Changing
the number of dimensions $D$ to evaluate loop integrals is well established
and relies on the analytic properties of loop integrals. We have used this
property to extend the number of dimensions to negative values as suggested
by Halliday and Ricotta. As discussed at length in Sec.~\ref{sec:general},
treating $D$ as a negative even integer allows a multinomial expansion of the
integrand in intermediate steps and, by expanding before and after loop
integration we can identify the loop integral as an infinite series, together
with constraints on the summation variables.  The number of summation
parameters is equal to the number of legs $n$ plus the number of energy
scales $(m+q)$ in the loop, while there are $n+1$ constraints, which can be
read off from the Feynman graph.  The form of the series is specified for
arbitrary one-loop integrals and forms a template series into which specific
solutions of the system of constraints are inserted. In this way, integration
over the parameters is replaced with infinite sums.  In each case we
immediately identify generalised hypergeometric functions and show how to
assemble the complete result valid in a particular kinematic region by
considering the convergence properties of the hypergeometric functions.  The
procedure is as follows:
\begin{enumerate}
\item
Write down $\P$, $\M$ and $\Q$ of Eqs.~(\ref{eq:P}), (\ref{eq:M})
and~(\ref{eq:Q}) by inspection of the graph.
\item
Write down the template solution for the particular process.   In other words,
copy out Eq.~(\ref{eq:sum}) inserting the correct mass scales and 
number of terms in $\P$, $\M$ and $\Q$.
\item
Construct the system of constraints by counting powers of $x_i$ in
Eqs.~(\ref{eq:left}) and (\ref{eq:right}).
\item
Solve the system of constraints and
insert each solution into the template solution, one at a time.
\item
Construct \poch\ symbols and identify the generalised hypergeometric
function.
\item 
Flip all of the $\Gamma$ functions in prefactor according to
Eq.~(\ref{eq:flip}).
\item
Group the solutions according to their regions of convergence.
\item 
Evaluate the hypergeometric functions for the specific parameters of interest.
\end{enumerate}
Steps 1-6 are very straightforward and easily achieved with a computer program.
Step 7 requires a little more thought, though for the cases we have studied
here the convergence regions were easy to identify.  To make the procedure
useful for phenomenological studies, it is necessary to evaluate the
hypergeometric functions for specific values of the parameters.   In
particular, an integral representation is required. The mathematical literature
for Euler integral representations of  hypergeometric functions with more than
2 or 3 variables is quite sparse, and it may be necessary to select a
complex integral representation for more complicated functions.

More interesting is the application of NDIM to integrals with more than one
loop. Suzuki and Schmidt \cite{SS2loop,SS3loop} have made some steps in this
direction, although the integrals they have considered are largely of the
one-loop insertion type.   Given that quite powerful results for one-loop
integrals are achieved so easily and generally, we expect that NDIM can play a
role in simplifying the task of calculating two- (or more) loop integrals
that are necessary to make more precise perturbative
predictions within the Standard Model.

\section*{Acknowledgements}
We thank J.V. Armitage for stimulating discussions and insights into the field
of hypergeometric functions and P. Watson and M. Zimmer for useful
conversations.   We acknowledge the assistance of D. Broadhurst and A.
Davydychev regarding generalised hypergeometric functions. C.A. acknowledges
the financial support of the Greek government and C.O.  acknowledges the
financial support of the INFN.

\appendix
%%%%%%%%%%%%%%%%%%%%%%%%%%%%%%%%%%%%%%%%%%%%%%%%% 
\section{Hypergeometric definitions and identities} 
\label{sec:app}
%%%%%%%%%%%%%%%%%%%%%%%%%%%%%%%%%%%%%%%%%%%%%%%%% 

The purpose of this appendix is to give sufficient information to evaluate
the general loop integrals presented in Secs.~\ref{sec:general}
and~\ref{sec:triangles}.  In Sec.~\ref{subsec:series} we give the definitions
of the hypergeometric functions as a series together with their regions of
convergence. Integral representations are provided in
Sec.~\ref{subsec:integral} together with a description of how to evaluate the
integrals in the general case. For specific choices of the $\nu_i$ the
general hypergeometric functions often simplify and some useful identities
and analytic-continuation formulae are collected in
Sec.~\ref{subsec:identities}.

%%%%%%%%%%%%%%%%%%%%%%%%%%%%%%%%%%%%%%%%%%%%%%%%% 
\subsection{Series representations} 
\label{subsec:series}
%%%%%%%%%%%%%%%%%%%%%%%%%%%%%%%%%%%%%%%%%%%%%%%%% 

The hypergeometric functions of one variable are sums of \poch\ symbols over
a single summation parameter $m$, like, for example,
\begin{eqnarray}
\label{eq:f21_def}
{_2F_1}\(\al, \bt, \ga,x\) 
&=& \sum_{m=0}^{\infty} \frac{(\al,m)(\bt,m)
}{(\ga,m)} \,\frac{x^m}{m!} 
\\
\label{eq:f32_def}
{_3F_2}\(\al, \bt, \bp ,\ga,\gp, x\) 
&=& \sum_{m=0}^{\infty} \frac{(\al,m)(\bt,m)(\bp,m)
}{(\ga,m)(\gp,m)} \,\frac{x^m}{m!}, 
\end{eqnarray}
which are convergent when $|x| < 1$.

We also meet hypergeometric functions of two variables which 
can be written as sums over the integers $m$ and $n$:  $F_i$, $i=1,\ldots,4$
are the Appell functions, $H_2$ a Horn function and $S_1$ and $S_2$ generalised
Kamp\'{e} de F\'{e}riet functions:
\begin{eqnarray}
\label{eq:f1_def}
F_1\(\al, \bt, \bp, \ga,x,y\) 
&=& \sum_{m,n=0}^{\infty} \frac{(\al,m+n)(\bt,m)(\bp,n)
}{(\ga,m+n)} \,\frac{x^m}{m!} \,\frac{y^n}{n!}
\\
\label{eq:f2_def}
F_2\(\al,\bt,\bp,\ga,\gp,x,y \) 
&=& \sum_{m,n=0}^{\infty} \frac{(\al,m+n)(\bt,m)(\bp,n)
}{(\ga,m)(\gp,n)} \,\frac{x^m}{m!}\, \frac{y^n}{n!}
\\
\label{eq:f3_def}
F_3\(\al,\ap,\bt,\bp,\ga,x,y \)
&=& \sum_{m,n=0}^{\infty} \frac{(\al,m)(\ap,n)(\bt,m)(\bp,n)
}{(\ga,m+n)} \, \frac{x^m}{m!}\, \frac{y^n}{n!} 
\\
\label{eq:f4_def}
F_4\(\al,\bt,\ga,\gp,x,y\) 
&=& \sum_{m,n=0}^{\infty} \frac{(\al,m+n)(\bt,m+n)}
{(\ga,m)(\gp,n)} \, \frac{x^m}{m!} \, \frac{y^n}{n!} 
\\
\label{eq:h2_def}
H_2\(\al,\bt,\ga,\gp,\de,x,y\) 
&=& \sum_{m,n=0}^{\infty} \frac{(\al,m-n)(\bt,m)(\ga,n)(\gp,n)}
{(\de,m)}\, \frac{x^m}{m!} \, \frac{y^n}{n!} 
\\
\label{eq:s1_def}
S_1\(\al,\ap,\bt,\ga,\de,x,y\)
&=& \sum_{m,n=0}^{\infty} \frac{(\al,m+n)(\ap,m+n)(\bt,m)}
{(\ga,m+n)(\de,m)}\, \frac{x^m}{m!} \, \frac{y^n}{n!} 
\\
\label{eq:s2_def}
S_2\(\al,\ap,\bt,\bp,\ga,x,y\)
&=& \sum_{m,n=0}^{\infty} \frac{(\al,m-n)(\ap,m-n)(\bt,n)(\bp,n)}{(\ga,m-n)}
\, \frac{x^m}{m!} \, \frac{y^n}{n!}. 
 \phantom{aaa}
\end{eqnarray}

These series converge according to the criteria collected in
Table~\ref{tab:convergence},
\begin{table}[ht]
\begin{center}
\begin{tabular}{c| c}
 \hline
Function  &  Convergence criteria \\
\hline
$F_1$, $F_3$ & $\displaystyle{|x| < 1}, \ |y| < 1$ \\
$F_2$, $S_1$ & $\displaystyle{|x| + |y| < 1}$ \\
$F_4$ & $\displaystyle{\sqrt{|x|} + \sqrt{|y|} < 1}$ \\
$H_2$ , $S_2$ & $ -|x| + 
%\displaystyle{\frac{1}{|y|}} 
1/|y| 
> 1$, $|x| <1$, $|y|<1$ \vspace{1mm}  \\
\hline
\end{tabular}
\caption{Convergence regions for some hypergeometric functions of two
variables.}
\label{tab:convergence}
\end{center}
\end{table}
The domain of convergence of the Appell and Horn functions are well known.  
That one for $S_1$ and $S_2$ may be worked out using Horns general
theory of convergence \cite{Exton}.

When one of the arguments vanishes, then the hypergeometric function
collapses in a straightforward way.  For example, if $y=0$ in
Eq.~(\ref{eq:s1_def}), then only the first term of the series in $n$
contributes and we are left with the relation,
\beq
\label{eq:S1_ytozero}
S_1\( \al,\ap,\bt,\ga,\de,x,0\) = {_3F_2}\(\al,\ap,\bt,\ga,\de,x\).
\eeq
Similarly, we have
\beqn
\label{eq:S1_xtozero}
S_1\(\al,\ap,\bt,\ga,\de,0,y\)  &=& \f21\(\al,\ap,\ga,y\) \\
S_2\( \al,\ap,\bt,\gp,\ga,x,0\) &=& \f21\(\al,\ap,\ga,x\) \\ 
\label{eq:S2_xtozero} 
S_2\(\al,\ap,\bt,\bp,\ga,0,y\) &=& {_3F_2}\(1-\ga,\bt,\bp,1-\al,1-\ap,-y\)\\
F_1\( \al,\bt,\bp,\ga,x,0\) &=& {_2F_1}\(\al,\bt,\ga,x\) \\ 
F_1\( \al,\bt,\bp,\ga,0,y\) &=& {_2F_1}\(\al,\bp,\ga,y\) \\ 
F_2\( \al,\bt,\bp,\ga,\gp,x,0\) &=& {_2F_1}\(\al,\bt,\ga,x\) \\ 
\label{eq:F2_xtozero}
F_2\(\al,\bt,\bp,\ga,\gp,0,y\)  &=&  \f21\(\al,\bp,\gp,y\) \\
F_3\( \al,\ap,\bt,\bp,\ga,x,0\) &=& {_2F_1}\(\al,\bt,\ga,x\) \\ 
F_3\( \al,\ap,\bt,\bp,\ga,0,y\) &=& {_2F_1}\(\ap,\bp,\ga,y\) \\ 
\label{eq:F4_ytozero}
F_4\( \al,\bt,\ga,\gp,x,0\) &=& {_2F_1}\(\al,\bt,\ga,x\) \\ 
\label{eq:F4_xtozero}
F_4\( \al,\bt,\ga,\gp,0,y\) &=& {_2F_1}\(\al,\bt,\gp,y\) \\ 
H_2\( \al,\bt,\ga,\gp,\de,x,0\) &=& {_2F_1}\(\al,\bt,\de,x\) \\ 
\label{eq:H2_xtozero}
H_2\( \al,\bt,\ga,\gp,\de,0,y\)  &=& \f21\(\ga,\gp,1-\al,-y\).
\eeqn

When one of the parameters vanishes producing a \poch\ $(0,n)$, then the
series in $n$ also terminates. If we have $(0,m+n)$ then both series
terminate. For example
\beqn
F_2\( \al,0,\bp,\ga,\gp,x,y\) &=& {_2F_1}\(\al,\bp,\gp,y\) \\ 
F_1\( 0,\bt,\bp,\ga,x,y\) &=& 1.
\eeqn

%%%%%%%%%%%%%%%%%%%%%%%%%%%%%%%%%%%%%%%%%%%%%%%%% 
\subsection{Integral representations} 
\label{subsec:integral}
%%%%%%%%%%%%%%%%%%%%%%%%%%%%%%%%%%%%%%%%%%%%%%%%% 

Euler integral representations of the hypergeometric series of one and two
variables are well known \cite{KdF,erdelyi} and we list them here.  We know
of no integral representation for the $H_2$ function and for the closely
related $S_2$ function.
\dl{
{_2F_1}\(\al,\bt,\ga,x\) = \frac{\G{\ga}}{\G{\bt}\G{\ga-\bt}}
\times \int_0^1 du  \, u^{\bt-1}
(1-u)^{\ga-\bt-1} (1-u x)^{-\al}
\hfill}
\beq
\label{eq:f21_integral}
\Re(\bt)>0, \quad \Re(\ga-\bt) >0.
\eeq
\dl{
{_3F_2}\(\al,\bt,\ga,\de,\ep,x\) = \frac{\G{\de}\G{\ep}}{\G{\bt}\G{\de-\bt}
\G{\ga} \G{\ep-\ga}}\hfill\cr
\phantom{{_3F_2}\(\al,\bt,\ga,\de,\ep,x\) =}
\times \int_0^1 du \int_0^1 dv \, u^{\bt-1} v^{\ga-1}
(1-u)^{\de-\bt-1} (1-v)^{\ep-\ga-1} (1-uvx)^{-\al}
\hfill
}
\beq
\label{eq:f32_integral}
\Re(\bt)>0, \quad \Re(\de-\bt) > 0, \quad \Re(\ga) > 0, \quad \Re(\ep-\ga)>0.
\eeq

\dl{
F_1(\al, \bt, \bp, \ga,x,y) 
= \frac{\G{\ga}}{\G{\al}\G{\ga-\al}} \int_0^1 du \, 
u^{\al-1}(1-u)^{\ga-\al-1}(1-ux)^{-\bt}(1-uy)^{-\bp}
}
\beq
\label{eq:f1_integral}
 \Re(\al) > 0, \quad \Re(\ga-\al) > 0.
\eeq
\dl{
F_2\(\al,\bt,\bp, \ga,\gp,x,y\) = \frac{\G{\ga}\G{\gp}}{\G{\bt}\G{\bp}
\G{\ga-\bt} \G{\gp-\bp}}\hfill\cr
\phantom{F_2\(\al,\bt,\bp, \ga,\gp,x,y\) =}
\times \int_0^1 du \int_0^1 dv \, u^{\bt-1} v^{\bp-1}
(1-u)^{\ga-\bt-1} (1-v)^{\gp-\bp-1} (1-ux-vy)^{-\al}\hfill
}
\beq
\label{eq:f2_integral}
\Re(\bt)>0, \quad \Re(\bp) > 0, \quad \Re(\ga-\bt) > 0, \quad \Re(\gp-\bp)>0.
\eeq
\dl{
F_3\(\al,\ap, \bt, \bp, \ga, x, y \) = \frac{\G{\ga}}{\G{\bt} \G{\bp}
\G{\ga-\bt-\bp}} \hfill\cr
\phantom{F_3\(\al,\ap, \bt, \bp, \ga, x, y \) = }
\times \int\!\! \int_{u \geq 0,\, v \geq 0 }^{u+v \leq 1} 
du\, dv\, u^{\bt-1} v^{\bp-1} (1-u-v)^{\ga-\bt-\bp-1}
(1-ux)^{-\al} (1-vy)^{-\ap}\hfill
}
\beq
\label{eq:f3_integral}
 \Re(\bt)>0, \quad \Re(\bp)>0, \quad \Re(\ga-\bt-\bp)>0.
\eeq
\dl{
F_4\(\al,\bt,\ga,\gp,x(1-y), y(1-x)\) =
\frac{\G{\ga}\G{\gp}} {\G{\al}\G{\bt}\G{\ga-\al} \G{\gp-\bt}}\hfill\cr
\hspace{5cm}
\mbox{}\times \int_0^1 du \int_0^1 dv \,
u^{\al-1} v^{\bt-1} (1-u)^{\ga-\al-1}(1-v)^{\gp-\bt-1} \hfill\cr
\hspace{5cm}
\mbox{}\times
(1-ux)^{\al-\ga-\gp+1} (1-vy)^{\bt-\ga-\gp+1}
(1-ux-vy)^{\ga+\gp-\al-\bp-1} \hfill
}
\beq
\label{eq:f4_integral}
\Re(\al) > 0, \quad \Re(\bt) > 0, \quad \Re(\ga-\al) >0, \quad
\Re(\gp-\bt) >0. 
\eeq

\dl{
S_1\(\al,\ap,\bt,\ga,\de,x,y\) = 
\frac{\G{\ga}}{\G{\al}\G{\ga-\al}} \int_0^1 du \, 
u^{\al-1}(1-u)^{\ga-\al-1} F_2(\ap,\bt,1,\de,1,ux,uy)
\hfill \cr
\phantom{S_1\(\al,\ap,\bt,\ga,\de,x,y\) = }
= \frac{\G{\ga}\G{\de}}{\G{\al}\G{\ga-\al}
\G{\bt} \G{\de-\bt}}\hfill\cr
\phantom{S_1\(\al,\ap,\bt,\ga,\de,x,y\) =}
\times \int_0^1 du \int_0^1 dv \, u^{\al-1} v^{\bt-1}
(1-u)^{\ga-\al-1} (1-v)^{\de-\bt-1} (1-uvx-uy)^{-\ap}\hfill
}
\beq
\label{eq:s1_integral}
\Re(\al)>0, \quad \Re(\ga-\al) > 0, \quad \Re(\bt) > 0, \quad \Re(\de-\bt)>0.
\eeq

%%%%%%%%%%%%%%%%%%%%%%%%%%%%%%%%%%%%%%%%%%%%%%%%% 
\subsubsection{Example of explicit evaluation of an integral representation}
\label{subsec:ep_expansion}
%%%%%%%%%%%%%%%%%%%%%%%%%%%%%%%%%%%%%%%%%%%%%%%%% 
In working out the integral representation for hypergeometric functions in
$D=4-2\ep$ dimensions, we have often to deal with the $\ep$ expansion of
integrals of the form
\beqn
\label{eq:def_Ix}
I(x)  &=& \int_0^1 du\, d(u)\, f(u),\\
  d(u) &=& u^{-1+\al\ep} (1-u)^{-1+\bt\ep} 
\eeqn
where $\al$ and $\bt$ are real numbers and $f(u)$ is a smooth function in
the domain $0 \le u \le 1$: in particular, it is finite at the boundary
points.  

The procedure to deal with this kind of integrals is quite standard.  The
integral has a 
pole in $\ep$ when the integration variable $u$ approaches either
of the end points.  We concentrate first on the point $u=0$, and we rewrite the
integral in such a way to expose the pole in $\ep$
\begin{equation}
\label{eq:fu_f0}
I(x)  = \int_0^1 du\, d(u)\,   f(0)  +
 \int_0^1 du\, d(u)\, 
\Big[ f(u) - f(0) \Big]  = I_{[1]}+I_{[2]}.
\end{equation}
The integral $I_{[1]}$ can be easily done
\begin{equation}
I_{[1]} = f(0) \frac{\G{\al\,\ep}\G{\bt\,\ep}}{\G{(\al+\bt)\,\ep}} =
\frac{f(0)}{\ep} \frac{\al+\bt}{\al\bt}   \frac{\G{1+\al\,\ep} \G{1+\bt\,\ep}}
{\G{1+(\al+\bt)\,\ep}},
\end{equation}
and the integrand of $I_{[2]}$ is now finite in the limit $u\,\to\,0$. In fact,
we can make a Taylor expansion  
\begin{equation}
   f(u) - f(0) = u f'(0) + \frac{u^2}{2!}f''(0) +\ldots \equiv u \, g(u),
\end{equation}
and write $I_{[2]}$ as
\begin{equation}
I_{[2]} = \int_0^1 du\, d(u)\, u \, g(u) = \int_0^1 du\,
 u^{\al\ep} (1-u)^{-1+\bt\ep}   g(u).
\end{equation}
We repeat now the same steps done for Eq.~(\ref{eq:fu_f0}) with respect to
the point $u=1$, to obtain
\begin{equation}
I_{[2]} = \int_0^1 du\, u^{\al\ep} (1-u)^{-1+\bt\ep} \, g(1) 
+ \int_0^1 du\, u^{\al\ep} (1-u)^{-1+\bt\ep} \,
\Big[g(u)-g(1) \Big]  =  I_{[3]}+I_{[4]}.
\end{equation}
The integral $I_{[3]}$ gives
\begin{equation}
I_{[3]} = g(1) \, \frac{\G{1+\al\,\ep} \G{\bt\,\ep}}{\G{1+(\al+\bt)\,\ep}} =
\frac{f(1)-f(0)}{\bt\,\ep} \,\frac{\G{1+\al\,\ep} \G{1+\bt\,\ep}}
{\G{1+(\al+\bt)\,\ep}},
\end{equation}
while  $I_{[4]}$ is finite at $u\,\to\,1$ 
\begin{equation}
I_{[4]} = \int_0^1 du\, u^{\al\ep} (1-u)^{\bt\ep} \, h(u), \qquad 
g(u)-g(1) \equiv (1-u) \, h(u),  
\end{equation}
and can be solved with an $\ep$ expansion of the integrand.
Adding all the contributions together we have
\begin{equation}
I(x) = \frac{1}{\al \, \bt\,\ep} \left[ \bt \, f(0) + \al \, f(1)  \right]
\frac{\G{1+\al\,\ep} \G{1+\bt\,\ep}} {\G{1+(\al+\bt)\,\ep}}
+ \int_0^1 du\, u^{\al\ep} (1-u)^{\bt\ep} \, h(u), 
\end{equation}
where
\begin{equation}
  h(u) = \frac{1}{u(1-u)} \biggl( f(u)-(1-u)\,f(0)-u\,f(1)\biggr).
\end{equation}

In the case where we have two integration variables, the procedure outlined
above can be re-iterated in a straightforward manner.  To illustrate the
procedure, we evaluate  explicitly the $F_2$ functions of
Eq.~(\ref{eq:I3m1p3}) to $\ord{\ep^2}$.\\ 
The integral representation for $F_2$ (see Eq.~(\ref{eq:f2_integral})) is
given by
\begin{equation}
F_2\(1,1,\ep,\ep+1,1-\ep,x,y\) = \frac{\ep^2\, \G{1-\ep}}{\G{1+\ep}
\G{1-2\ep}} \,\, I(x,y),
\end{equation}
where
\begin{equation}
\label{eq:def_I}
I(x,y)  = \int_0^1 du\, dv\,  d(u,v)\, f(u,v),
\end{equation}
and
\beqn
  d(u,v) &=& v^{-1+\ep} (1-u)^{-1+\ep} (1-v)^{-2\ep} \nonumber\\
  f(u,v) &=& (1-ux-vy)^{-1}, \nonumber
\eeqn
and $I(x,y)$ must be computed to $\ord{\ep^0}$.
In order to expose the poles (see Eq.~(\ref{eq:fu_f0})), we add and subtract
the value of the finite function $f(u,v)$, computed at the boundary points,
in the following way:
\beqn
 I(x,y) &=& 
\int_0^1 du\, dv\, d(u,v)\, \Big\{ \Big[f(1,0)\Big] + \Big[ f(u,0) - f(1,0)
\Big] + \Big[ f(1,v) -f(1,0) \Big] 
\nonumber\\
&& \hspace{4cm}+ \Big[ f(u,v) -f(u,0) -f(1,v) +f(1,0)
\Big]  \Big\} 
\nonumber \\
&=& I_{[1]}+I_{[2]}+I_{[3]}+I_{[4]}.
\eeqn
We are now in a position to evaluate the single contributions in the square
brackets.
In fact
\beqn
I_{[1]} &=&  (1-x)^{-1} 
 \int_0^1 du\,  (1-u)^{-1+\ep} \int_0^1 dv\, v^{-1+\ep}\,
 (1-v)^{-2\ep}= (1-x)^{-1}  \frac{\G{1+\ep} \G{1-2\ep}}{\ep^2\, \G{1-\ep}}
\nonumber\\
I_{[2]} &=& \frac{-x}{1-x}
\frac{\G{1+\ep} \G{1-2\ep}}{\ep \, \G{1-\ep}}
 \int_0^1 du\, \frac{(1-u)^\ep}{(1-ux)}\nonumber\\
I_{[3]} &=& \frac{(1-x)^{-1} }{\ep}
\int_0^1 dv\,  \frac{ v^{\ep} \,(1-v)^{-2\ep}}{ 1-x-vy}\nonumber\\
I_{[4]} &=& \frac{x y}{1-x} \int_0^1 du\,dv\, 
(1-u)^\ep \,v^{\ep} \,(1-v)^{-2\ep}\frac{(v y+u x + x-2)}
{ (1-u x)(1-x-v y)(1-v y-u x)}.
\eeqn
The remaining integrals are finite in the limit $\ep\,\to\,0$, so that we can
make a Taylor expansion to $\ord{\ep}$ for the integrands of $I_{[2]}$
and $I_{[3]}$, 
and we can put directly $\ep=0$ in $I_{[4]}$.  
Recalling the definition of the dilogarithm function
\begin{equation}
\label{eq:def_li2}
\li{x} = -\int_0^x dz \, \frac{\log(1-z)}{z} \hspace{1cm} x \leq 1\;,
\end{equation}
it is straightforward to carry on the last integrations and express the
result in terms of Li$_2$ functions, as done in Eq.~(\ref{eq:I3_ep_expans}).

%%%%%%%%%%%%%%%%%%%%%%%%%%%%%%%%%%%%%%%%%%%%%%%%% 
\subsection{Identities amongst the hypergeometric functions} 
\label{subsec:identities}
%%%%%%%%%%%%%%%%%%%%%%%%%%%%%%%%%%%%%%%%%%%%%%%%% 

There are three kinds of identities that relate hypergeometric functions.
First there are analytic continuations which connect functions in different
regions of convergence.  Second are reduction formula which allow the
functions to be expressed as simpler series for certain values of the
parameters. Finally there are transformations which relate the same functions
with different arguments.

\subsubsection{Analytic continuation formulae}

Here we give only those analytic continuation properties that relate the
argument and inverse argument.
Gauss' hypergeometric function has the following analytic continuation
properties (see for example \cite{erdelyi})
\dl{
{_2F_1}\(\al,\bt,\ga,z\) = 
(-z)^{-\al}
\frac{\G{\ga}\G{\bt-\al}}{\G{\bt}\G{\ga-\al}}
\, {_2F_1}\(\al,1+\al-\ga,1+\al-\bt,\frac{1}{z}\) \hfill\cr
\phantom{{_2F_1}\(\al,\bt,\ga,z\) =}
+ (-z)^{-\bt}
\frac{\G{\ga}\G{\al-\bt}}{\G{\al}\G{\ga-\bt}}
\, {_2F_1}\(\bt,1+\bt-\ga,1+\bt-\al,\frac{1}{z}\).
\hfill}
\beq
\label{eq:f21_anal_1/z}
| {\rm arg}(-z)| < \pi,
\eeq
\dl{
{_2F_1}\(\al,\bt,\ga,z\) = 
z^{-\al}\frac{\G{\ga}\G{\ga-\al-\bt}}{\G{\ga-\al}\G{\ga-\bt}}
\, {_2F_1}\(\al,1+\al-\ga,1+\al+\bt-\ga,1-\frac{1}{z}\) \hfill\cr
%\phantom{{_2F_1}\(\al,\bt,\ga,z\) =}
\hfill
+ z^{\al-\ga}\(1-z\)^{\ga-\al-\bt}
\frac{\G{\ga}\G{\al+\bt-\ga}}{\G{\al}\G{\bt}}
\, {_2F_1}\(\ga-\al,1-\al,1+\ga-\al-\bt,1-\frac{1}{z}\)
%\hfill
}
\beq
\label{eq:f21_anal_1-1/z}
| {\rm arg}(z)| < \pi, \  | {\rm arg}(1-z)| < \pi.
\eeq

The corresponding analytic continuation of the hypergeometric functions with
two variables are summarised in Table~\ref{tab:analcont}.  There are many
possible analytic continuations; however, we list only those that are
relevant to link the groups of solutions for the two-mass triangle integral
discussed in Sec.~\ref{subsec:2mass}, that is the connections between the
Appell and Horn functions. The others are easily derived by summing the
series with respect to one of the summation variables to obtain an ${_2F_1}$,
applying Eq.~(\ref{eq:f21_anal_1/z}), rewriting Gauss' hypergeometric
function as a series and reidentifying the double series.  We see that these
functions appear to form a group.

\begin{table}[ht]
\begin{center}
\begin{tabular}{c| c}
 \hline
Function  &  Continued in terms of \\
\hline
$F_4(x,y)$ & $F_4(x/y,1/y)$, $F_4(y/x,1/x)$\\
$F_3(x,y)$ & $H_2(1/x,-y)$, $H_2(1/y,-x)$, $F_2(1/x,1/y)$\\
$H_2(x,y)$ & $F_2(x,-1/y)$ \\
$F_2(x,y)$ & $S_1(-y/x,1/x)$, $H_2(y,-1/x)$, $S_1(-x/y,1/y)$, $H_2(x,-1/y)$\\
$H_2(x,y)$ & $F_3(1/x,-y)$, $S_2(1/x,-xy)$ \\
$S_1(x,y)$ & $F_2(-x/y,1/y)$ \\
$S_2(x,y)$ & $H_2(1/x,-xy)$ \\
\hline
\end{tabular}
\caption{Analytic continuation for the hypergeometric functions of two
variables.}
\label{tab:analcont}
\end{center}
\end{table}

\beqn
\label{eq:f4_anal_cont}
F_4\!\(\al,\bt,\ga,\gp,x, y\) \!\!\!\! &=& \!\!\!\! 
\frac{\G{\gp}\G{\bt-\al}}{\G{\gp-\al}\G{\bt}} (-y)^{-\al}
F_4\!\(\al,\al+1-\gp,\ga,\al+1-\bt,\frac{x}{y},\frac{1}{y}\) \nonumber\\
&+&\!\!\!\!
\frac{\G{\gp} \G{\al-\bt}}{\G{\gp-\bt} \G{\al}} (-y)^{-\bt}
F_4\!\(\bt,\bt+1-\gp,\ga,\bt+1-\al,\frac{x}{y},\frac{1}{y}\)\\
&&\nonumber \\
\label{eq:f3_anal_cont}
F_3\!\(\al,\ap,\bt,\bp,\ga,x,y\)\!\!\!\! &=& \!\!\!\!
\frac{\G{\bt-\al}\G{\ga}}{\G{\ga-\al}\G{\bt}}
(-x)^{-\al} H_2\!\(\al+1-\ga,\al,\ap,\bp,\al+1-\bt,\frac{1}{x},-y\) \nonumber
\\ 
&+&\!\!\!\!
\frac{\G{\al-\bt}\G{\ga}}{\G{\ga-\bt}\G{\al}}
(-x)^{-\bt} H_2\!\(\bt+1-\ga,\bt,\ap,\bp,\bt+1-\al,\frac{1}{x},-y\) \nonumber \\
&&\\
\label{eq:h2_anal_cont2}
H_2\!\(\al,\bt,\ga,\gp,\de,x,y\)\!\!\!\! &=& \!\!\!\! 
\frac{\G{\gp-\ga}\G{1-\al}}{\G{1-\al-\ga}\G{\gp}}
(y)^{-\ga} F_2\!\(\al+\ga,\bt,\ga,\de,\ga+1-\gp,x,-\frac{1}{y}\) \nonumber \\
&+&\!\!\!\!
\frac{\G{\ga-\gp}\G{1-\al}}{\G{1-\al-\gp}\G{\ga}}
(y)^{-\gp} F_2\!\(\al+\gp,\bt,\gp,\de,\gp+1-\ga,x,-\frac{1}{y}\) \nonumber \\
&&\\
\label{eq:f2_anal_cont}
F_2\!\(\al,\bt,\bp,\ga,\gp,x,y\)\!\!\!\! &=& \!\!\!\!
\frac{\G{\bt-\al}\G{\ga}}{\G{\ga-\al}\G{\bt}}
(-x)^{-\al} S_1\!\(\al,\al+1-\ga,\bp,\al+1-\bt,\gp,-\frac{y}{x},\frac{1}{x}\) \nonumber \\
&+&\!\!\!\!
\label{eq:h2_anal_cont1}
\frac{\G{\al-\bt}\G{\ga}}{\G{\ga-\bt}\G{\al}}
(-x)^{-\bt} H_2\!\(\al-\bt,\bp,\bt,\bt+1-\ga,\gp,y,-\frac{1}{x}\) \\
&&\nonumber \\
H_2\!\(\al,\bt,\ga,\gp,\de,x,y\)\!\!\!\! &=& \!\!\!\! 
\frac{\G{\bt-\al}\G{\de}}{\G{\de-\al}\G{\bt}}
(-x)^{-\al} S_2\!\(\al,\al+1-\de,\gp,\ga,\al+1-\bt,\frac{1}{x},-xy\) \nonumber \\
&+&\!\!\!\!
\frac{\G{\al-\bt}\G{\de}}{\G{\de-\bt}\G{\al}}
(-x)^{-\bt} F_3\!\(\bt,\gp,\bt+1-\de,\ga,\bt+1-\al,\frac{1}{x},-y\) \nonumber \\
&&\\
\label{eq:s1_anal_cont}
S_1\!\(\al,\ap,\bt,\ga,\de,x,y\)\!\!\!\! &=& \!\!\!\!
\frac{\G{\ap-\al}\G{\ga}}{\G{\ga-\al}\G{\ap}}
(-y)^{-\al} F_2\!\(\al,\bt,\al+1-\ga,\de,\al+1-\ap,-\frac{x}{y},\frac{1}{y}\) \nonumber \\
&+&\!\!\!\!
\frac{\G{\al-\ap}\G{\ga}}{\G{\ga-\ap}\G{\al}}
(-y)^{-\ap} F_2\!\(\ap,\bt,\ap+1-\ga,\de,\ap+1-\al,-\frac{x}{y},\frac{1}{y}\) \nonumber \\
&&\\
\label{eq:s2_anal_cont}
S_2\!\(\al,\ap,\bt,\bp,\ga,x,y\)\!\!\!\! &=& \!\!\!\!
\frac{\G{\ap-\al}\G{\ga}}{\G{\ga-\al}\G{\ap}}
(-x)^{-\al} H_2\!\(\al,\al+1-\ga,\bp,\bt,\al+1-\ap,\frac{1}{x},-xy\) \nonumber \\
&+&\!\!\!\!
\frac{\G{\al-\ap}\G{\ga}}{\G{\ga-\ap}\G{\al}}
(-x)^{-\ap} H_2\!\(\ap,\ap+1-\ga,\bp,\bt,\ap+1-\al,\frac{1}{x},-xy\) \nonumber \\
\eeqn

\subsubsection{Reduction formulae}
\label{app:reduction}
The $F_4$ functions describing the massive bubble and the off-shell massless
triangle have the following reduction formulae which leave a single remaining
Euler integral at most \cite{erdelyi,bailey}
\beqn
&&\!\!\!F_4\(\al,\bt,\ga,\bt,-\frac{x}{(1-x)(1-y)},-\frac{y}{(1-x)(1-y)}\) 
\nonumber\\
&&\!\!\!\hspace{5cm} = (1-x)^{\al}(1-y)^{\al}
F_1\(\al,\ga-\bt,1+\al-\ga,\ga,x,xy\), \\ 
&&\!\!\!F_4\(\al,\bt,\al,\bt,-\frac{x}{(1-x)(1-y)},-\frac{y}{(1-x)(1-y)}\) =
(1-xy)^{-1} (1-x)^{\bt} (1-y)^{\al}, \phantom{aaaaaa} \\
&&\!\!\!F_4\(\al,\bt,\bt,\bt,-\frac{x}{(1-x)(1-y)},-\frac{y}{(1-x)(1-y)}\)
\nonumber\\ 
&&\!\!\!\hspace{5cm}  =(1-x)^{\al}(1-y)^{\al} \f21\( \al,1+\al-\bt,\bt,xy\), \\
&&\!\!\!F_4\(\al,\bt,1+\al-\bt,\bt,-\frac{x}{(1-x)(1-y)},-\frac{y}{(1-x)(1-y)}
\)   
\nonumber\\ 
&&\!\!\!\hspace{5cm}  = (1-y)^{\al}
\f21\(\al,\bt,1+\al-\bt,-\frac{x(1-y)}{1-x} \). 
\eeqn
Similar reductions for the other functions of two variables are
\beqn
&&F_1\( \al,\bt,\bp,\bt+\bp,x,y \) = 
(1-y)^{-\al}\f21\(\al,\bt,\bt+\bp, \frac{x-y}{1-y} \) \\
&&F_2\(\al,\bt,\bp,\ga,\al,x,y\) = (1-y)^{-\bp}
F_1\(\bt,\al-\bp,\bp,\ga,x,\frac{x}{1-y} \) \\
&&F_2\(\al,\bt,\bp,\al,\gp,x,y\) = (1-x)^{-\bt}
F_1\(\bp,\bt,\al-\bt,\gp,\frac{y}{1-x},y\) \\
&&F_2\( \al,\bt,\bp,\bt,\gp,x,y\) = (1-x)^{-\al} \f21\(\al,\bp,\gp,
\frac{y}{1-x}\) \\
&&F_2\( \al,\bt,\bp,\al,\al,x,y\) = (1-x)^{-\bt} (1-y)^{-\bp}
\f21\(\bt,\bp,\al,\frac{xy}{(1-x)(1-y)}\)\phantom{aaaa}\\
&&F_2\( \al,\bt,\bp,\al,\bp,x,y\) = (1-y)^{\bt-\al} (1-x-y)^{-\bt}
\\
&&F_2\( \al,\bt,\bp,\bt,\bp,x,y\) = (1-x-y)^{-\al}\\
&&F_3\(\al,\ga-\al,\bt,\bp,\ga,x,y\) = (1-y)^{-\bp}
F_1\(\al,\bt,\bp,\ga,x,\frac{y}{y-1} \) \\
&&F_3\(\al,\ga-\al,\bt,\ga-\bt,\ga,x,y\) = (1-y)^{\al+\bt-\ga}
\f21\(\al,\bt,\ga,x+y-xy\),
\eeqn
while for certain values of the parameters the $H_2$ function reduces to an
$F_2$ or $F_1$
\beqn
H_2\(\al,\bt,\ga,\de-\al,\de,x,y\) &=& 
(1-x)^{-\bt} F_2\(\de-\al,\bt,\ga,\de,1-\al, -\frac{x}{1-x},-y\),
\phantom{aaa}\\ 
H_2\(\al,\bt,\ga,1-\al,\de,x,y\) &=& 
(1+y)^{-\ga} F_1\(\bt,\al,\ga,\de, x,\frac{xy}{1+y}\).
\eeqn

The $S_1$ and $S_2$ functions we have introduced are less well known.  
From the
integral representation or manipulating the series using Eq.~(\ref{eq:revhyp})
we find
\beqn
S_1\(\al,\ap,\bt,\al,\de,x,y\) &=& 
(1-y)^{-\ap} \f21\(\ap,\bt,\de,\frac{x}{1-y}\),\\
S_1\(\al,\ap,\bt,\ga,\bt,x,y\) &=& 
 \f21\(\al,\ap,\ga,x+y\), \\
S_2\(\al,\ap,\bt,\bp,\al,x,y\) &=& 
(1-x)^{-\ap} \f21\(\bt,\bp,1-\ap,-y(1-x)\).  
\eeqn

\subsubsection{Transformation formulae}

A useful formula connecting Gauss' hypergeometric function to itself is
\beqn
{_2F_1}\(\al,\bt,\ga,z\)  &=& 
\( 1-z \) ^{-\al} 
{_2F_1}\(\al,\ga-\bt,\ga,\frac{z}{z-1}\) \nonumber \\
&=&\( 1-z \) ^{-\bt} 
{_2F_1}\(\ga-\al,\bt,\ga,\frac{z}{z-1}\) \nonumber \\
\label{eq:revhyp}
&=&\( 1-z \) ^{\ga-\al-\bt} 
{_2F_1}\(\ga-\al,\ga-\bt,\ga,z\).
\eeqn 
Using the above results, then
\beqn
F_1\( \al,\bt,\bp,\ga,x,y\) 
&=& (1-x)^{-\bt}(1-y)^{-\bp}
    F_1\(\ga-\al,\bt,\bp,\ga,\frac{x}{x-1},\frac{y}{y-1} \) \nonumber \\
&=& 
(1-x)^{-\al} F_1\(\al,\ga-\bt-\bp, \bp,\ga,\frac{x}{x-1},\frac{x-y}{x-1}\) \nonumber \\
&=&
(1-y)^{-\al} F_1\(\al,\bt,\ga-\bt-\bp,\ga,\frac{y-x}{y-1},\frac{y}{y-1}\) \\
&& \nonumber \\
%%%%%%%%%%%%%%%%%%%%%
F_2\(\al,\bt,\bp,\ga,\gp,x,y \)
&=&(1-x)^{-\al} F_2\(\al,\ga-\bt,\bp,\ga,\gp,\frac{x}{x-1},\frac{y}{1-x}\) \nonumber \\
&=&(1-y)^{-\al} F_2\(\al,\bt,\gp-\bp,\ga,\gp,\frac{x}{1-y},\frac{y}{y-1}\) \nonumber \\
&=&(1-x-y)^{-\al}
F_2\(\al,\ga-\bt,\gp-\bp,\ga,\gp,\frac{x}{x+y-1},\frac{y}{x+y-1}\). \nonumber \\
%%%%%%%%%%%%%%%%%%%%%
\eeqn

%\clearpage

%%%%%%%%%%%%%%%%%%%%%%%%%%%%%%%%%%%%%%%%%%%%%%%%%%%%%%%%%%%%%%%%%%
%%%%%%%%%%%%%%%%%%%%    FOR NIGEL TO USE!!!!  %%%%%%%%%%%%%%%%%%%%
%%%%%%%%%%%%%%%%%%%%%%%%%%%%%%%%%%%%%%%%%%%%%%%%%%%%%%%%%%%%%%%%%%

\relax
\def\pl#1#2#3{{\it Phys.\ Lett.\ }{\bf #1}\ (19#2)\ #3}
\def\zp#1#2#3{{\it Z.\ Phys.\ }{\bf #1}\ (19#2)\ #3}
\def\prl#1#2#3{{\it Phys.\ Rev.\ Lett.\ }{\bf #1}\ (19#2)\ #3}
\def\rmp#1#2#3{{\it Rev.\ Mod.\ Phys.\ }{\bf#1}\ (19#2)\ #3}
\def\prep#1#2#3{{\it Phys.\ Rep.\ }{\bf #1}\ (19#2)\ #3}
\def\pr#1#2#3{{\it Phys.\ Rev.\ }{\bf #1}\ (19#2)\ #3}
\def\np#1#2#3{{\it Nucl.\ Phys.\ }{\bf #1}\ (19#2)\ #3}
\def\sjnp#1#2#3{{\it Sov.\ J.\ Nucl.\ Phys.\ }{\bf #1}\ (19#2)\ #3}
\def\app#1#2#3{{\it Acta Phys.\ Polon.\ }{\bf #1}\ (19#2)\ #3}
\def\jmp#1#2#3{{\it J.\ Math.\ Phys.\ }{\bf #1}\ (19#2)\ #3}
\def\jp#1#2#3{{\it J.\ Phys.\ }{\bf #1}\ (19#2)\ #3}
\def\nc#1#2#3{{\it Nuovo Cim.\ }{\bf #1}\ (19#2)\ #3}
\def\lnc#1#2#3{{\it Lett.\ Nuovo Cim.\ }{\bf #1}\ (19#2)\ #3}
\def\ptp#1#2#3{{\it Progr. Theor. Phys.\ }{\bf #1}\ (19#2)\ #3}
\def\tmf#1#2#3{{\it Teor.\ Mat.\ Fiz.\ }{\bf #1}\ (19#2)\ #3}
\def\tmp#1#2#3{{\it Theor.\ Math.\ Phys.\ }{\bf #1}\ (19#2)\ #3}
\def\jhep#1#2#3{{\it J.\ High\ Energy\ Phys.\ }{\bf #1}\ (19#2)\ #3}
\def\epj#1#2#3{{\it Eur.\ Phys. J.\ }{\bf #1}\ (19#2)\ #3}
\relax

\end{document}